\documentclass{egpubl}
\usepackage{eurovis2018}
\Full_EuroVis
\ConferencePaper        % uncomment for (final) Conference Paper
 \electronicVersion % can be used both for the printed and electronic version

\ifpdf \usepackage[pdftex]{graphicx} \pdfcompresslevel=9
\else \usepackage[dvips]{graphicx} \fi

\PrintedOrElectronic

\usepackage{t1enc,dfadobe}

\usepackage{egweblnk}
\usepackage{cite}

\graphicspath{{figures/}{pictures/}{images/}{./}} % where to search for the images
\usepackage{booktabs}
\usepackage[moderate]{savetrees}
\usepackage{balance}

\DeclareRobustCommand{\bfparhead}[1]{\noindent\textbf{#1.}} % CD: boldface \parhead  

\title[Track Xplorer]
      {Track Xplorer: A System for\\ Visual Analysis of Sensor-based Motor Activity Predictions}

      \author[Marco Cavallo \& \c{C}a\u{g}atay Demiralp]
{\parbox{\textwidth}{\centering Marco Cavallo$^{1}$
        and \c{C}a\u{g}atay Demiralp$^{1}$ 
        }
        \\
{\parbox{\textwidth}{\centering $^1$IBM Research}
}
}

\begin{document}

\maketitle

\begin{abstract}

With the rapid commoditization of wearable sensors, detecting human movements
from sensor datasets has become increasingly common over a wide range of
applications. To detect activities, data scientists iteratively experiment with
different classifiers before deciding which model to deploy. Effective
reasoning about and comparison of alternative classifiers are crucial in
successful model development.  This is, however, inherently difficult in
developing classifiers for sensor data, where the intricacy of long temporal
sequences, high prediction frequency, and imprecise labeling make standard
evaluation methods relatively ineffective and even misleading.

We introduce Track Xplorer, an interactive visualization system to query,
analyze, and compare the predictions of sensor-data classifiers. Track Xplorer
enables users to interactively explore and compare the results of different
classifiers, and assess their accuracy with respect to the ground-truth labels
and video. Through integration with a version control system, Track Xplorer
supports tracking of models and their parameters without additional workload on
model developers. Track Xplorer also contributes an extensible algebra over 
track representations to filter, compose, and compare classification outputs, 
enabling users to reason effectively about classifier performance.   
 
We apply Track Xplorer in a collaborative project to develop classifiers to
detect movements from multisensor data gathered from Parkinson's disease
patients.  We demonstrate how Track Xplorer helps identify early on possible
systemic data errors, effectively track and compare the results of different
classifiers, and reason about and pinpoint the causes of
misclassifications.   
%\begin{classification} % according to http://www.acm.org/class/1998/
  %\CCScat{Human-centered computing}{I.3.3}{Human computer interaction
  %(HCI)}{Interactive systems and tools} \end{classification}
\end{abstract}

%-------------------------------------------------------------------------
\section{Introduction}\label{sec:intro}

\begin{figure}[tb]
     \centering
     \includegraphics[width=\columnwidth]{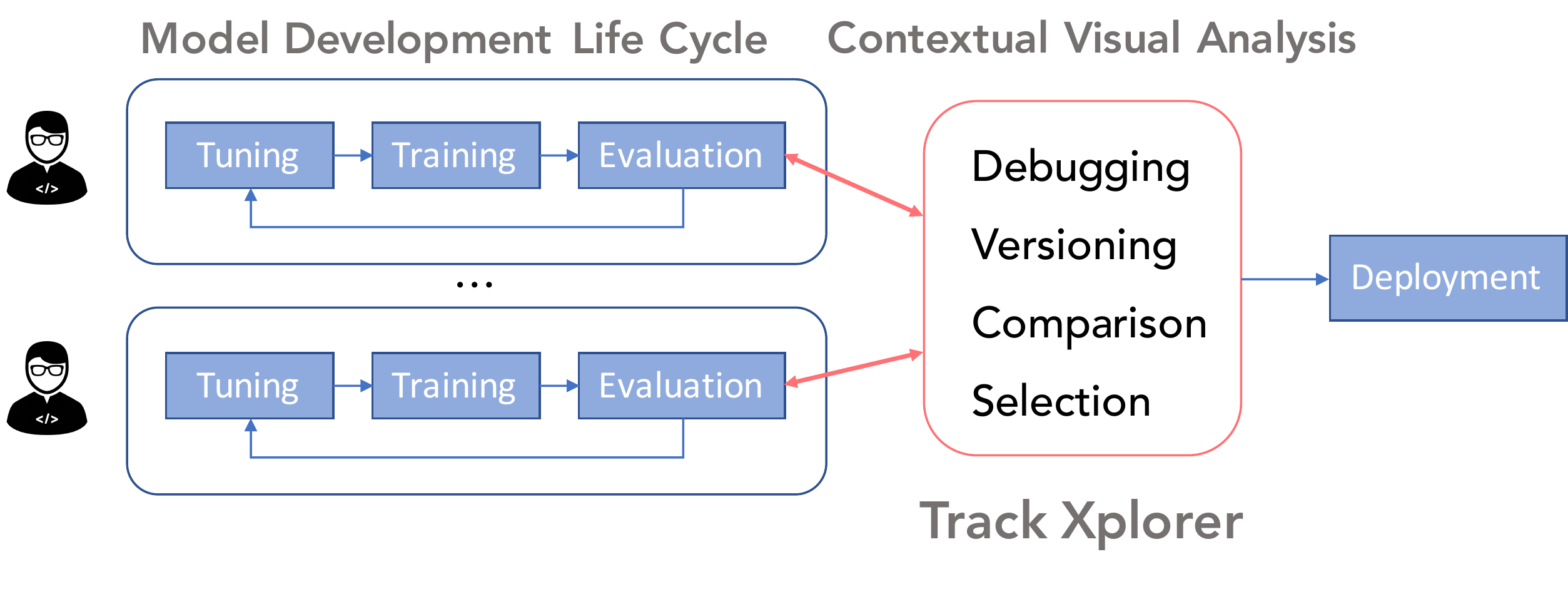}
     \caption{Improving model development life cycle. Track Xplorer 
       enhances the comparative evaluation of multiple predictive 
       models through contextual visual analysis.\label{fig:lifecycle}} 
\end{figure}
    
\begin{figure*}
    {
      \centering
      \includegraphics[width=\linewidth]{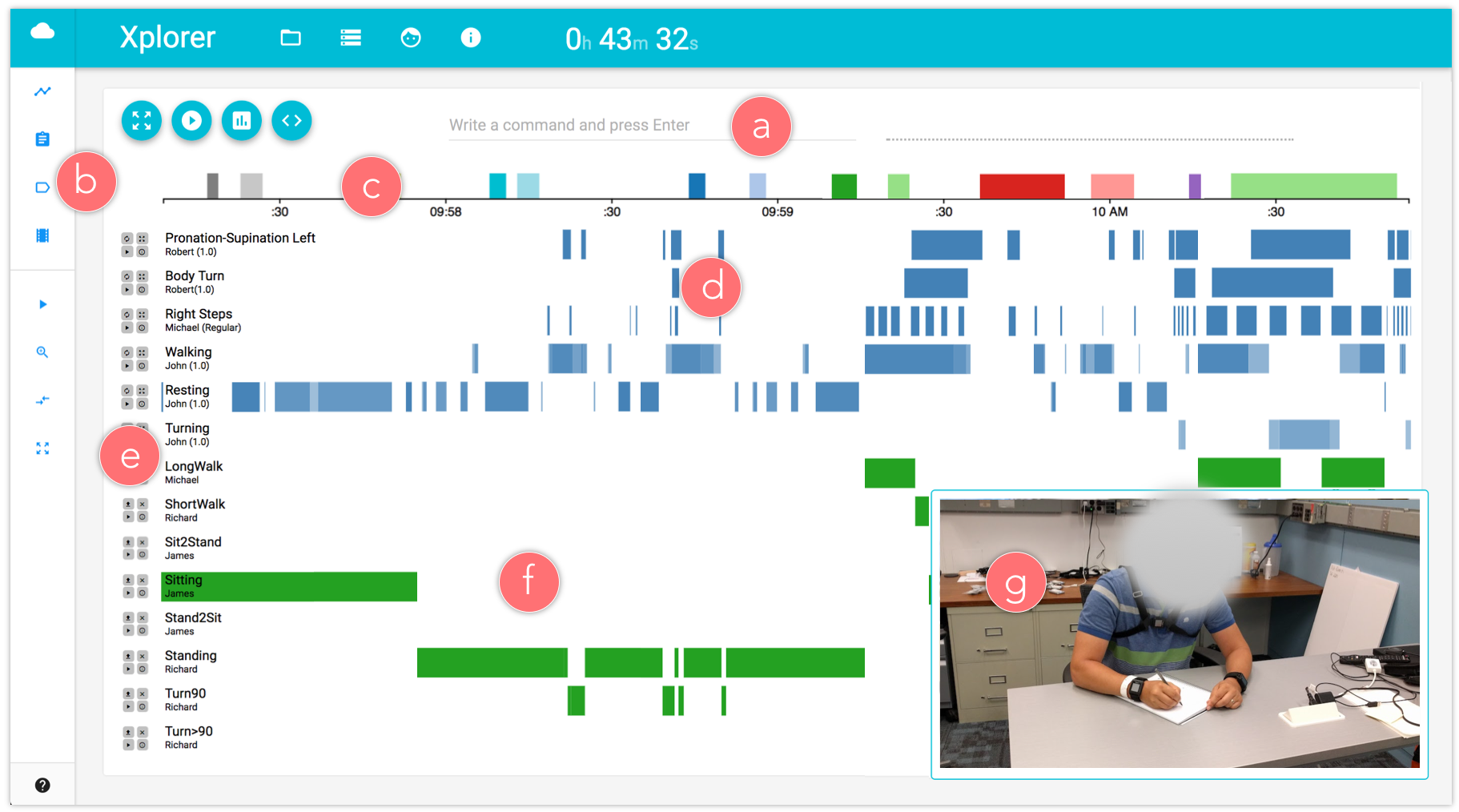}
      \caption{Track Xplorer interface. The main view shows a timeline for a
	set of stacked linear tracks. A track can represent the sequence of
	event predictions generated by a classifier (d, in blue) or the series
	of annotations used as ground-truth labels (f, in green).  Tracks are
	associated to the predicted motor activity, to the name of the person
	who generated the results or labels and to the model version (in case
	of classifiers). This information is displayed at the left margin of
	each track (e), together with four buttons that enable performing track
	operations. A dedicated protocol track (c) can be used to easily
	navigate to regions of interest, while a synchronized video player (g)
	lets users validate the context of event predictions. Users can
	customize the appearance of tracks through a collapsible sidebar on the
	left (b). A command-line interface (a) lets users run commands from 
	Track Xplorer's command set, which also supports the operations of a
	visual track algebra.
	\label{fig:teaser}} 
     }
\end{figure*}

The extensive diffusion of consumer-level wearable devices has opened up
many possibilities for activity monitoring and analysis. Fitness trackers,
from smart watches to wristbands, are increasingly used by people to track 
their daily activities. Similarly, a wide variety of biosensors is starting
to play an important role in continuous outpatient monitoring. Detecting
motor activities (i.e., limb movements) from sensor data and understanding
their normal and abnormal variation has a great potential to inform and
predict the well-being of healthy persons and patients alike. To this end,
developing classifiers that successfully detect different types of physical
motion from sensor data is critical for the reliability of insights derived
from any downstream analysis. 

Classifier development is an iterative process in which data scientists
start with some alternative models and associated hyperparameters, train
the models on a subset of the available data, and then evaluate their
performance on test data. During the evaluation step, data scientists use
aggregate performance metrics such as accuracy score, precision and recall
to establish how well a classifier predicts specific activities.  Data
scientists then tune their algorithms and parameters using these insights,
restarting the cycle and iteratively improving their models.  Analyzing the
output of predictive models is a crucial step in the life cycle of
iterative model development (Fig.~\ref{fig:lifecycle}). Although aggregate
evaluation metrics can be useful for overall performance assessment and
reporting, they don't carry sufficient detail to facilitate
interpretability and drill-down analysis needed to debug and
fine-tune the models. Aggregate metrics treat predictions uniformly, hiding
differences within them. Researchers have recently proposed interactive
tools (e.g.,~\cite{modeltracker2015chi,wheel2014tvcg,squares2017tvcg}) to
help evaluate classifier performances both in aggregate and at instance
level, primarily using interactive histograms of instance class
probabilities.

When evaluating motor activity classifiers, the inadequacy of aggregate
metrics becomes even more pronounced due to the intricate temporal nature of
data that classifiers operate on.  Each activity and its time window is
recognized by considering multiple model predictions computed at high frequency, and
is often validated against labels generated through manual video
annotation. The need to evaluate a classifier's output against manual 
labels that are generally affected by human interpretation, goal 
and bias further limits the effectiveness of aggregate performance metrics
to convey insights into \textit{why} one prediction model appears
to perform better than another. Data scientists need tools that would enable
them to drill down to the context of each motor activity prediction to better
interpret the causes of mispredictions. Existing tools using histograms of
instance class probabilities are, however, insufficient for the visual
analysis of activity classifications from sensor-data streams, as they do not consider
the temporal context of predictions. 

We introduce a novel visualization system, \textit{Track Xplorer}
(\textit{Xplorer} for short), to interactively analyze the results of
sensor-data classifiers. Xplorer enables users to debug and compare multiple
classifiers down to the granularity of a single prediction (instance),
providing qualitative and quantitative means to validate the performance of
each model. Xplorer facilitates the interpretation of classification results
in application context and thus lets data scientists reason about the causes
of misclassifications and improve their predictive models. 
%We note that our
%system does not aim at gaining insights on the internal behavior of a
%predictive model; instead, its purpose is to analyze a model's output.

\bfparhead{Contributions} Our contributions include (1) a novel
visualization system for visual analysis of classification performances on
temporal data, (2) a set of methods to integrate contextual information and
model metadata to support performance analysis, and (3) a visual track
algebra for filtering, composing, analyzing and comparing classification
results.  To illustrate the usefulness of our system, we present a use case
in developing predictive models to detect specific motor activities in
individuals with Parkinson's disease.  We study the use of Xplorer through
a group of fourteen participants, data scientists and business managers,
working on the same project.  We demonstrate that Xplorer proved essential
in validating and comparing predictive models, reasoning about the
causes of mispredictions, and understanding the trade-offs in the usage of
different sensors---improving the overall predictive-model-development life
cycle. We further observe how the system facilitates discussion among data 
scientists and business managers.  

We first give a synopsis of prior work, followed by a brief discussion of
our system design. We then provide details on Track Xplorer's interactions
and visual design along with its track algebra, command-line set and
classification validation support.  Next we discuss the use of Track
Xplorer in developing classifiers to detect movement patterns in
Parkinson's disease. We conclude by summarizing our contributions 
and offering a take-home.

\section{Related Work}\label{sec:related}
Our work is related to earlier research in systems infrastructure for improving
the machine-learning (ML) model development cycle, interactive analysis of
classifier performances, sequential and temporal data query and visualization,
and systems that facilitate visual analysis through algebraic operations.
    
\subsection{Systems Infrastructure for Improving ML Pipeline}
The practical success of ML in general and deep learning in particular has
dramatically increased the demand for applying ML models to solve problems
across domains. However, ML model development is iterative and time-consuming
and often relies on trial and error using extensive engineering skills and
large training data that are expensive and difficult to acquire and
maintain~\cite{sculley2015hidden}.

Recent database research proposes systems to improve the ML-modeling life
cycle.  ModelDB~\cite{vartak2016modeldb} stores ML models along with
associated pipelines and parameters, enabling the exploration of these models
through SQL queries. Xplorer's visual track algebra specializes in
comparative analysis of model predictions and performance and can be
integrated into a SQL-like declarative language.  
ModelHub~\cite{miao2017modelhub} provides a custom model-versioning system to track models and a
domain-specific language to configure deep networks and their
hyperparameters.  Xplorer supports model tracking through integration with
\textit{git}, which is already used by developers to track the source code of
their models. This avoids  the burden and fragmentation of using two separate
version control systems. Xplorer automatically extracts the metadata (e.g.,
parameters, training data info, etc.) for a given model along with its
version from the model's source code in the git repository.

Training-data collection and management are also critical in end-to-end
machine learning. Snorkel~\cite{ratner2017snorkel} supports training-data
generation at scale for ML models using weak supervision through user-defined
labeling functions~\cite{ratner2016data}. DataHub~\cite{datahub} supports
managing datasets and their versions over time. Van der Weide et
al.~\cite{van2017versioning} tie data versioning to ML pipelines, tracking
data resulting from intermediate stages of ML pipelines to reduce the
redundancy of computations and improve their robustness.
ProvDB~\cite{miao2017provdb} tracks the provenance of artifacts (e.g., data,
scripts, results) from data science experiments using git and a graph
database. Similar to VisTrails~\cite{callahan2006vistrails}, ModelDB, ProvDB,
and \cite{van2017versioning} also use workflows (pipelines) to represent
modeling or analysis processes. However, unlike VisTrails, where users
manually design the workflows, these three tools generate or infer workflows
for users.  Xplorer is akin to ProvDB~\cite{miao2017provdb} in passively
collecting model metadata. Unlike workflow-based systems, Xplorer focuses on
a specific step in the ML pipeline: analysis of model prediction and
performance. It is not intended to author and manage workflows for ML models.
In this sense, Xplorer can be considered as a microservice that can be plugged into
ML workflow management systems.   

\subsection{Visual Analysis of Classifier Performance}

Researchers have introduced interactive tools,
e.g.,~\cite{squares2017tvcg,modeltracker2015chi,wheel2014tvcg}, to help
data scientists make sense of their classifiers' performances.
Squares~\cite{squares2017tvcg} supplements summary performance statistics
with instance-level distribution information to uncover distinct
characteristics of classifiers with comparable aggregate performance.
Similarly, ModelTracker~\cite{modeltracker2015chi} and Confusion
Wheel~\cite{wheel2014tvcg} tightly couple performance with data instances
to enable multiscale analysis.  

These earlier tools combine histogram visualizations of instance class
probabilities with aggregate metrics. However, the histogram visualizations
of earlier work are not sufficient for drill-down analysis of
sensor-data classifications, as they hide the temporal context of the
classifier predictions. Track Xplorer lets users effectively inspect,
compare, combine and reason about instance classifications while
effectively visualizing the temporal context of these classifications.
Xplorer also complements earlier work on classifier performance analysis by
focusing on temporal data classifications, integrating  additional ``human
soft knowledge'' (e.g., activity videos and expert labels) and introducing
a visual algebra over classification results that enables composable
and rigorous performance analysis.   

\subsection{Sequential Data Visualization and Querying}

The visual design of Xplorer is drawn from genomic data browsers (e.g.
\cite{ucscbrowser2002gr,igv:natureb11,ensembl2004gr,savant:bioinf10}) and
multimedia editors~\cite{imovie,premiere}, using visual encoding along a
linear axis (track) of data and metadata sequences as the basic unit of
representation. Genomic browsers enable the visualization of molecular
sequences from various sources as aligned linear tracks that  can be added,
removed and reordered on demand. Genomic browsers support interactions such
as zooming and panning to enable fine-grained exploration of the data,
often encoded as variable-length horizontal bars. These features are also
common in multimedia editors, where tracks typically represent audio or
video sources, and are shared by many other tools in the temporal and
sequential data visualization
literature~\cite{lifelines1996chi,lifeflow2011chi,timelines1994gi,
karam1994sigsoft,bade2004chi}. 

% Discuss earlier work 
To query temporal and sequential data effectively, prior research also
proposes using a new temporal logic~\cite{allen1983cam}, SQL
extensions~\cite{jensen1999review,snodgrass1994tsql}, regular
expressions~\cite{zgraggen2015chi,cappers2018exploring}, graphical
languages~\cite{fails2006vast,monroe2013chi}, and visual analogy with comic
strips~\cite{querymarvel2009vlhcc, querymarvel2010vast}.
DecisionFlow~\cite{gotz2014decisionflow} introduces a milestone-based
analytics for event sequences, whereas EventPad~\cite{cappers2018exploring}
applies a regular-expression-based visual language for identifying patterns
of interest.  Monroe et al.'s  work ~\cite{monroe2013chi} focuses instead
on the specification of overlap and absence of interval events.  

The primary use of our algebra is to generate new meaningful 
temporal sequences from the combination of existing ones through  
various operators, instead of matching event sequences of interest. 
Thus, our visual track algebra over temporal data-classification results
complements earlier work on query-based selection. 
%The track algebra is applied in the domain of sensor data classification,
%where time intervals are dependent on classifier parameters and continuous
%probabilities.

%Track algebra
\subsection{Visual Analysis and Design Through Algebraic Operations} 

Earlier data visualization work proposes algebraic operators over data as
well as visual encoding and design variables to formulate and compose
complex hypotheses and visualizations.  Polaris~\cite{Stolte2002}
introduces a table algebra drawn from Wilkinson's grammar of
graphics~\cite{Wilkinson_2005}. Vega-Lite~\cite{vegalite2017infovis} uses
a composition algebra to construct layered and small-multiple views of
visualizations. invis~\cite{Demiralp2013} provides an algebraic approach to
inspecting RNA sequences represented as linear tracks, where mutations can
be visually aggregated using the logical operators AND, OR, and NOT. Track
Xplorer builds on earlier work and introduces an algebra that facilitates
effectively manipulating, combining, and comparing track representations of
classification results.

\section{Track Xplorer}\label{sec:design}
% ---- DESIGN OF XPLORER ---- 

   \begin{figure}[tb]
     \centering
     \includegraphics[width=\columnwidth]{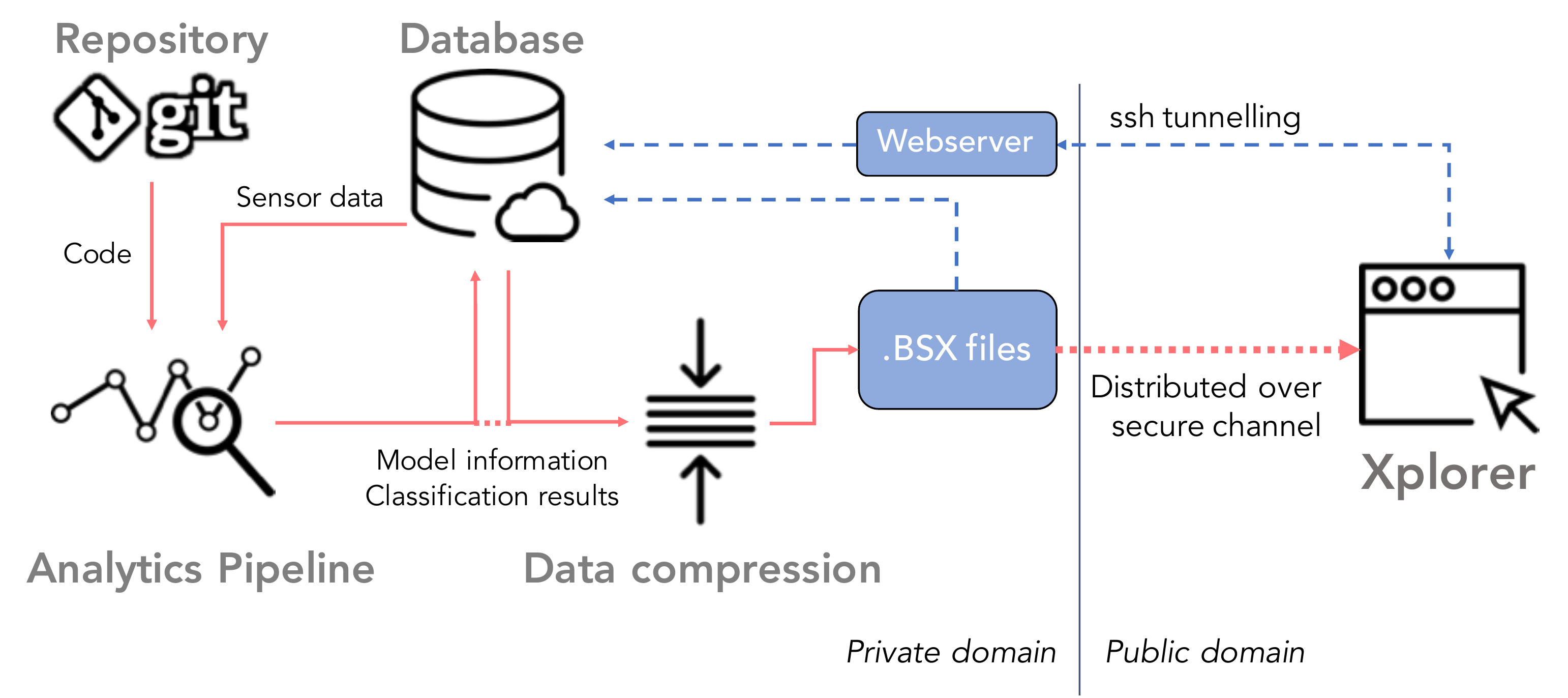}
     \caption{System design. Our analytics pipeline automatically 
       loads models from the code repository and generates classification 
       results for sensor data, storing the output in the database.
       Classification results and model metadata are then exported 
       as a compressed (.BSX) file that Xplorer uses as input. Xplorer optionally supports direct access to the database through ssh tunneling (shown with blue arrows in the figure above), avoiding the need for sharing .BSX files.       \label{fig:system}}
     \end{figure}
    
    \begin{figure}[tb]
     \centering
     \includegraphics[width=\columnwidth]{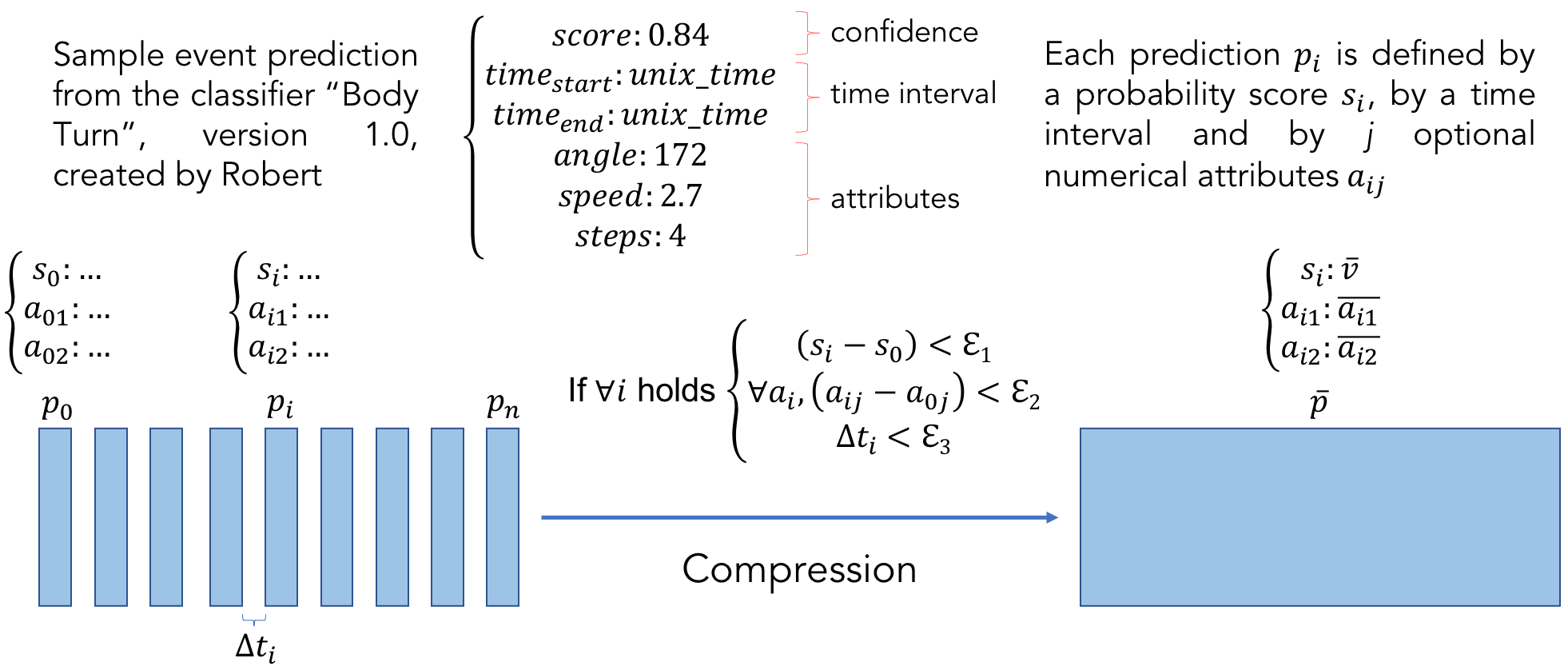}
     \caption{
       Data compression. Prediction results adjacent to each other in time and
       meeting specific similarity criteria can be aggregated into a single
       longer prediction. For each classifier prediction $p_i$, $s_i$ denotes its confidence value, $a_{ij}$ its attribute values and $\Delta t$ its time distance from the end of the previous prediction. Compression is performed if these values do not vary beyond certain thresholds $\epsilon_k$ from the first prediction. \label{fig:compression}}
     \end{figure}

\subsection{Design Considerations}
%The problem
Interacting with activity predictions based on sensor data poses a wide set
of design challenges related to temporal and computational scalability. 
Running predictive models on large amounts of sensor data can take 
several hours and, depending on the sampling rate, can generate 
many predictions over time, making interactive analysis 
difficult. This is particularly noticeable when classifiers output continuous probability values over time through a sliding-window approach, which can generate multiple prediction samples per second.
    
%What we do
Our system tackles these issues through automated precomputation and data
compression. In particular, we rely on a set of procedures that
periodically handle accessing the available predictive models, 
running them on the right subset of the sensor data, and compressing and
storing their output (Fig.~\ref{fig:system}). Data scientists are required to include in their code a standardized set of attributes that hold
meta-information about the nature, version and requirements of a
classifier. This information is complemented by the \textit{git} version
control system, which holds data about user commits and modifications in
the code. An analytics pipeline checks the code repository for new versions
of the models and runs them when a change is detected,
performing the computation according to the meta-information extracted from
each classifier. Results are then stored in a centralized database together
with the classification results. By merging similar predictions close to
each other in time (Fig.~\ref{fig:compression}), we compress the results
and store them in a JSON-based file (.BSX) that can later be opened from
the public, web-based Xplorer user interface.

%Distribution and confidentiality
An important design criterion for our system involves preserving
data confidentiality while allowing easy distribution of classification
results. Since data science projects are often run across groups or companies, it is common to maintain separate access privileges for prediction
results, sensor data and source code. Distributing BSX files means that
any user can easily be given access to prediction results through the
publicly available Xplorer interface, without being exposed to any confidential
information. On the other hand, data scientists can easily enable an optional,
protected ssh tunneling connection (Fig.~\ref{fig:system}) to access
confidential information without having to deal with BSX files.

%Generalizability
While we here propose a very specific system architecture to optimize the overall model development cycle, Xplorer can also be used as a standalone tool to handle the import of classifier predictions, ground-truth labels or generic time intervals from standard CSV files.

%Additional stuff
%We refer readers to Supplementary Material for further implementation details.

    \begin{figure}[tb]
     \centering
     \includegraphics[width=\columnwidth]{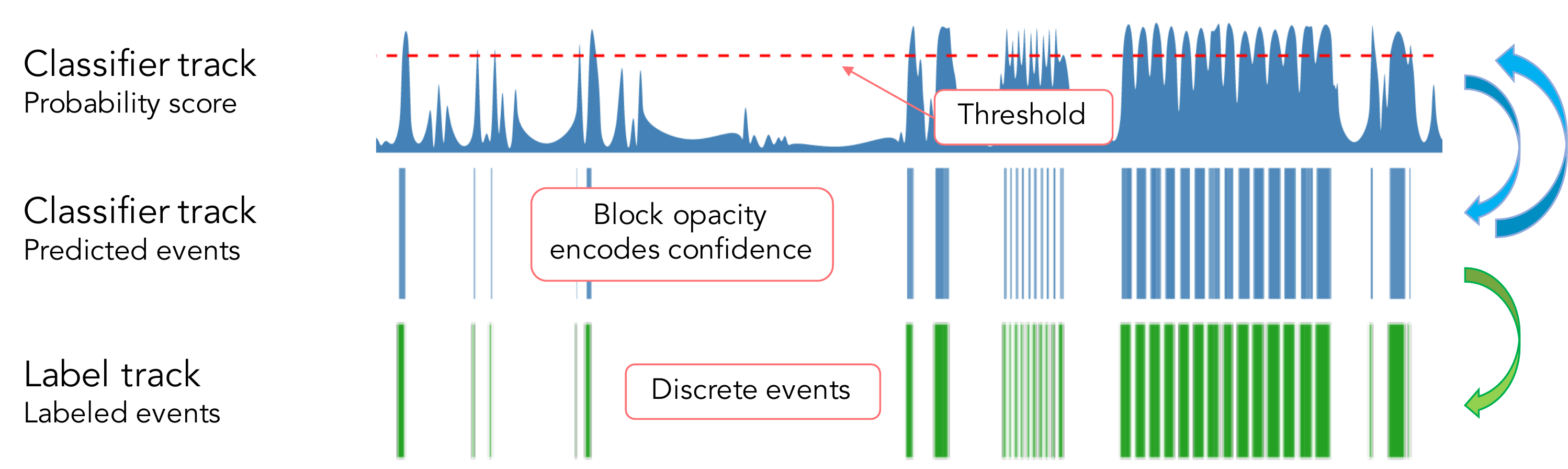}
     \caption{
       Track definition. A track in Xplorer corresponds to a list of non-overlapping time periods
       (``events''). The two track types are based on the form of the
       data they represent: classifier tracks and label tracks. A
       classifier track contains probability scores associated to each event
       and can be visualized either as an area chart or as horizontal bars
       (``blocks''), whereas a label track contains only information about time
       intervals. A classifier track can be converted into a label track by
       applying a threshold on the classifier track's prediction scores.\label{fig:track_types}}
    \end{figure}

%-------------------------------------------------------------------------

\subsection{User Interface}
The interface of Xplorer (Fig.~\ref{fig:teaser}) is composed of a main
    view in which classification results and labels are represented as linear
    tracks stacked vertically. A track visually corresponds to a sequence of
    non-overlapping colored blocks positioned over a common timeline. We
    categorize tracks into two types based on the form of the data they
    visualize: classifier tracks and label tracks (Fig.~\ref{fig:track_types}). Although here we represent classifier tracks in blue and label tracks in green, we note that both the color of each track and their vertical arrangement are freely customized by the user.
    
    %Classifier tracks
    In classifier tracks (Fig.~\ref{fig:teaser}a), a block
    corresponds to a single prediction or to a set of consecutive 
    identical predictions, and this may result in blocks of variable length.  If the output
    of a classifier is binary, the block is visible only when the activity is
    detected.  If a classifier outputs continuous probability scores, the block
    is generated by applying a classifier-specific threshold to the score
    values (Fig.~\ref{fig:track_types}). The opacity of a block encodes the
    associated probability score, increasing with high score values and
    decreasing with low score values. When dealing with multi-class classifiers,
    Xplorer generates by default a separate track for each predicted class,
    leaving an option to aggregate them visually.   
    
    %Label tracks
    In label tracks (Fig.~\ref{fig:teaser}b), each block corresponds to a
    textual label (e.g. ``Walking,'' ``Person is sitting'') characterized by a
    start and end time that determine its position and length. Labels can be
    either algorithmically generated or manually defined by a human, and are
    often used as ground-truth or as a reference for validating classifier
    tracks. 
    % 
    % The
    % color of blocks in a label track, differently than in classifier tracks, has
    % always the same opacity.
    %Protocol
    A particular type of label track, called \textit{protocol track}
    (Fig.~\ref{fig:teaser}c), can hold different unique labels on the same
    timeline, given they do not overlap with each other. A protocol track is
    generally used as a reference to the data collection process, where each 
    block corresponds to a specific task performed by a subject.  
    
    While all other temporal data is represented in linear tracks, video is
    shown in a separate undocked window (Fig.~\ref{fig:teaser}d), which can be
    dragged across the interface and freely resized by the user.  The interface
    of Xplorer also includes a left sidebar from which users can decide which
    tracks to visualize and can easily zoom to specific events in the
    protocol track, and three auxiliary modal windows. These latter windows can be used
    for (1) analyzing classifier performance, (2) inspecting model information
    and code revisions, and (3) accessing raw sensor data associated to a
    particular prediction or time interval.
    
    \begin{figure}[tb]
     \centering
     \includegraphics[width=\columnwidth]{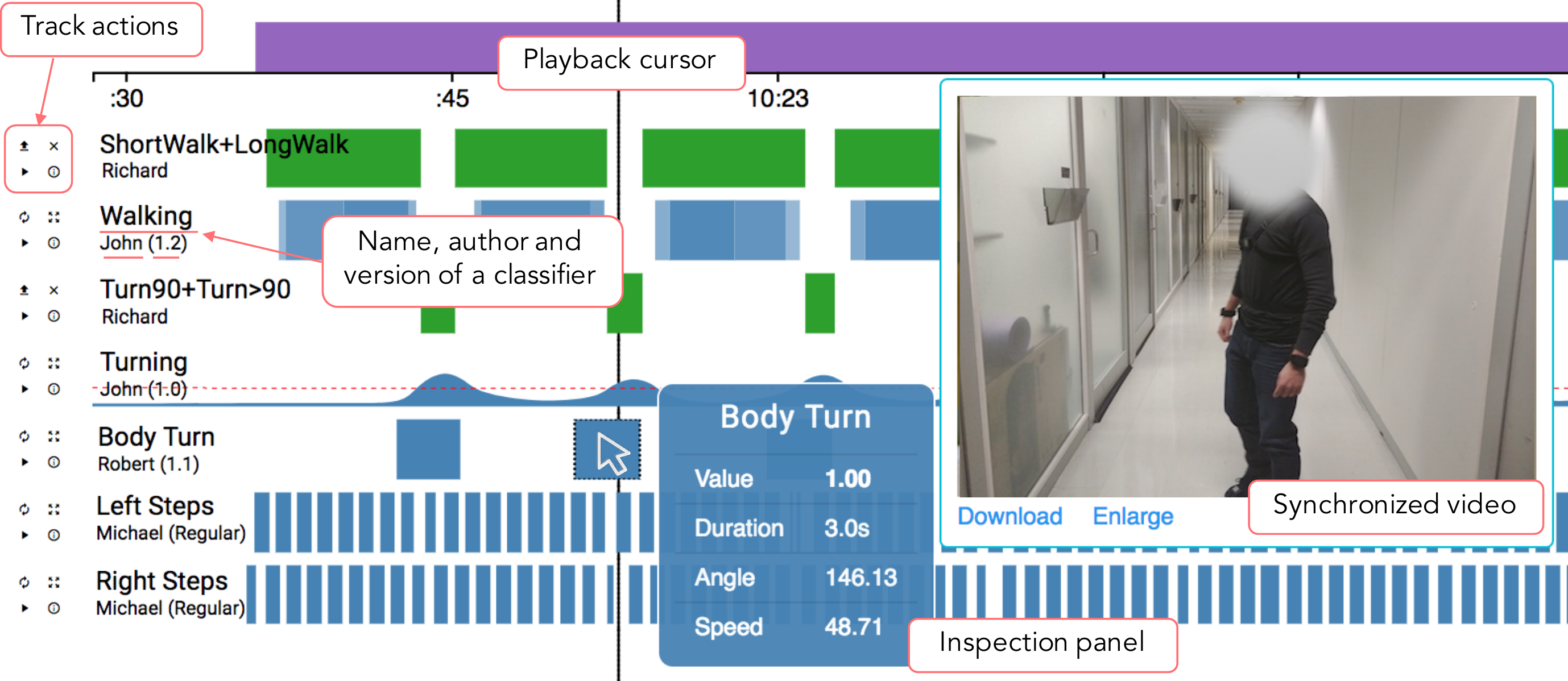}
     \caption{Sample track configuration in Xplorer. It is obvious from the pattern of
       green and blue boxes (label and classifier tracks, respectively) that the subject is alternating between walking and turning movements. Visual alignment, video playback and variable
       inspection can be used in combination to better understand predicted
       motor events.
     %Sample tracks from BlueSky Project. By observing the pattern of
     %green boxes (label tracks), it is straightforward to observe that the subject
     %is alternating walking to turning movements. 
     %While the alignment of the
     %classifier tracks ``Walking'' and ``Turning'' with their correspondent
     %ground-truth labels seems satisfactory, we may note that the predictive model
     %``Body Turn'' appears to detect events too early.
     %By inspecting sensors usage
     %in the video and the attributes associated to each prediction (shown as a
     %tooltip), data scientists can try to debug their prediction models. The
     %possibility to dynamically change the threshold of a classifier (as shown on
     %the ``Turning'' track) further aims at a better understanding of predicted
     %motor events.
     \label{fig:detail}}
    \end{figure}
    
    \subsection{Interactions}
    Track information can be analyzed at different levels of granularity
    through zooming and panning, which are performed with the mouse wheel and
    drag actions.
    % In order to maintain the time alignment among predictions and labels,
    % each movement transform is applied equally to all tracks. 
    %By double clicking on a block or by selecting the name of a protocol
    %label from the left sidebar, all tracks are conveniently scaled to show
    %a close-up of the moment of interest.
    Hovering on a block displays information about the correspondent
    prediction or label (e.g. author, duration) as a tooltip. For classifier
    tracks, the tooltip shows classifier-specific variables associated to the
    prediction (e.g. ``tremor frequency,'' ``angular velocity'') that data
    scientists can use to debug their algorithms (Fig.~\ref{fig:detail}).
    
    Each classifier track also includes four buttons enabling the user to (1)
    increase its height for better visibility, (2) play consecutively the
    videos of all detected activities, (3) display information about the
    underlying predictive model (e.g. sensors, prediction window and
    threshold used), and (4) switch between two different visualization modes
    (area chart or horizontal bars).  Fig.~\ref{fig:track_types} explains how
    a classifier track can also be represented as an area chart, visualizing
    a continuous probability score over time. This mode is particularly
    useful for observing how a classifier's threshold determines which events
    are detected (and thus generate a block). The threshold can be
    dynamically changed by the user by moving the red horizontal line shown
    in Fig.~\ref{fig:track_types}, thus avoiding the recomputation of
    classification results.  All tracks can be vertically rearranged by
    dragging, so that users can better compare them by placing tracks of
    interest next to each other.
    
    %(PE) Finally, users can select a time range through brushing. This can
    %be used as a convenience function to export the raw sensor data related
    %associated to that time interval or to restrict the process of
    %validation (presented later in this paper) to a subset of the data.

\section{Visual Track Algebra}\label{sec:trackalgebra}
% --- VISUAL TRACK ALGEBRA ---

 \begin{figure}[tb]
     \centering  
     \includegraphics[width=\columnwidth]{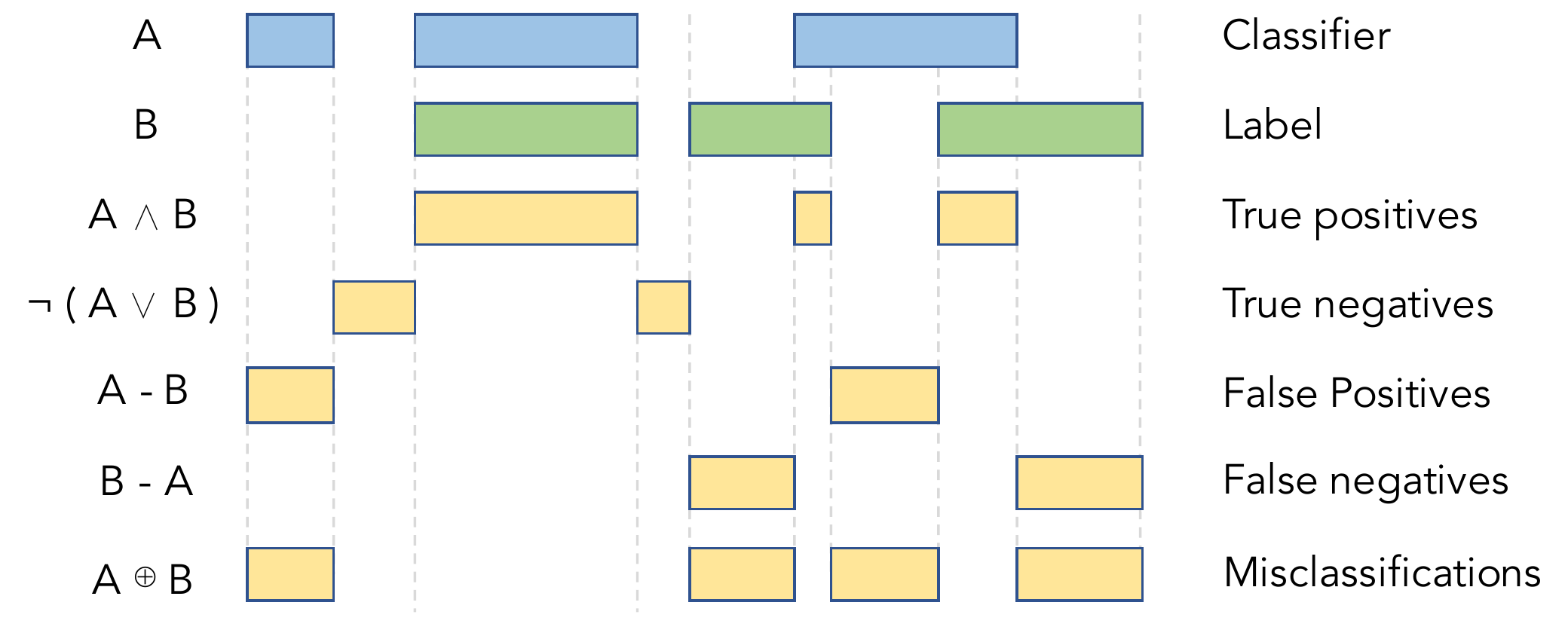}
     \caption{
     Validation of classifiers through track algebra. $A$ represents a
     classifier track and $B$ represents a label track containing ground-truth
     labels. By computing the intersection and subtraction of the two tracks,
     users can quickly identify correct and incorrect
     predictions.\label{fig:track_algebra}}
    \end{figure}

While analyzing each track separately may be sufficient for some applications,
in many cases the ability to combine different tracks can be essential.  For
instance, a user may want to analyze the output of a tremor classifier only
when a different classifier predicts no walking movement. Similarly, a user
wanting to consider all moments in which a subject is stationary needs to unify
the labels associated to ``Sitting'' with labels associated to ``Standing''.
To enable reasoning beyond the scope of single classifiers and labels, we
define a visual algebra that can generate new tracks as combinations of
existing tracks. Operations such as addition, subtraction, logic conjunction
and disjunction can be applied to both classifier and label tracks with
different semantic meaning.
    
    Fig.~\ref{fig:track_algebra} illustrates how the most common operators can be
    used for classifier validation.  If we denote a classifier track by $A$ and a
    label track used as ground truth by $B$, $A \wedge B$ corresponds to the
    intersection of both tracks, that is to the events that were correctly
    predicted by the model (true positives). Similarly, we can define the difference
    between track $A$ and track $B$ as a new track composed of instances of $A$ not present in $B$. This way, the track $A-B$ contains all
    classifications that do not match any ground-truth label (false positives),
    while $B-A$ conversely represents labeled events not identified
    by the predictive model (false negatives).
    
    The power of the track algebra lies in enabling users quickly to combine
    tracks to validate complex hypotheses about the classification process. In
    particular, in the presence of ground-truth labels, identifying
    misclassified events is visually straightforward. In combination with the video
    functionality, the track algebra also enables the consecutive playing of all false positive and 
    false negative predictions for a particular classifier. This way, the user can
    visually validate the performance of his predictive model and reason about the
    causes of each single misprediction based on its context.

    % \begin{table}[]
    \begin{figure}
    \centering
    \resizebox{\linewidth}{!}{
    \begin{tabular}{rccc}
    \toprule
    Operator & $P_1$ & $P_2$ & Description                   \\ \midrule
    negate &         $T$      &        & Generates $\neg\;T$                           \\
    add  / union        &       $T_1$      &      $T_2$       & Generates $T_1\; \vee\; T_2$ \\
    intersection &         $T_1$      &      $T_2$       & Generates $T_1\; \wedge\; T_2$                           \\
    errors &         $T_1$      &      $T_2$       & Generates $T_1\; \oplus\; T_2$                           \\
    subtract     &       $T_1$       &        $T_2$      & Generates $T_1\; -\; T_2$ \\
     match     &       $T_1$       &        $T_2$      & Generates $T_1\; \sim\; T_2$ \\
    variation         &      $T_1$       &    [$T_2$]  & Generates $T_1\; \partial\;\; T_2$ \\                 
    
    play         &       $T$      &             & Plays all events in track $T$                            \\
    threshold           &     $C$        &     Float        & Changes C's threshold to a fixed value                            \\
    show / hide           &      $T$       &             & Shows / hides track T
    \\
    transform          &       $C$     &     & Thresholds $C$ and generates an $L$ track \\
    rename         &      $T$       &             & Renames a track  \\
    color         &      $T$       &   Color          & Changes the color of track T  \\
    author         &      $A$       &             & Shows author A's tracks \\
        filter         &      $T$       &  Exp           & Selects events of T that satisfy $Exp$  \\   
        order         &             &             & Smart ordering based on current time	\\
        info         &        $T$     &             & Display detailed information on a track	\\
    jaccard          &       $T_1$      &     $T_2$        & Jaccard distance between $T_1$ and $T_2$ \\
    roc          &       $C$      &     $L$        & Computes ROC curve and AUC score \\
    report &  $C$      &     $L$  & Computes precision, recall and F1 score  \\
    score         &      $T$       &             & Computes the match score metric     
        
    %goto         &      $E$       &             & Zooms all tracks to a particular event    \\
    \\ \bottomrule
    \end{tabular}}
    \caption{Commands of the Xplorer command line. $P_1$ and  $P_2$ are
      the parameters required by each command. $T$ is a placeholder for a generic
      track's identifier, whereas $C$ and $L$ indicate a classifier and a label
      track respectively. Track type conversion is handled automatically according
      to Fig.~\ref{fig:track_types}.\label{tab:commands}}
    \end{figure}

\subsection{Command Line: Combining, Filtering and Ordering}
    Track Xplorer interface features a command-line interface that lets users 
    quickly perform complex interactions, such as track manipulation through visual
    algebra. Fig.~\ref{tab:commands} lists the most common commands 
    executable from the command line. Each command is composed of one operator
    and one or two operands, which can be track identifiers or numerical values. A
    track identifier is automatically generated as a combination of the track name,
    author and version (e.g., the first version of the ``Sleeping'' classifier
    created by author ``John'' generates the ID ``SleepingJohn1.0'') and is
    made available by auto-completion.
    %Highlighting
    For instance, when ``threshold sleep'' is typed, the command line 
    automatically infers which available track is best suited to the operator
    ``threshold'', highlights it in the main view, and offers the suggested
    completion ``threshold SleepingJohn1.0''. Tracks can also be referred by their order of appearance (e.g. ``union 1 2'' generates the union of the first two tracks) or by string wildcards (e.g. ``show \%walk'' makes all tracks related to walk visible).
	%Generating a new track
    When a command generates a new track, it is added to the main view and
    its name and identifier are defined automatically from the operation
    performed.
    The command line can also be used to compute performance metrics, order tracks, and filter classification events based on their attributes. For instance, the command ``filter TurningErhan1.2 angle>60\&duration>2'' creates a new track containing only slow-turn events that last more than two seconds and in which the subject rotates by more than sixty degrees. When a moment $t$ along the timeline is chosen with the cursor, the command ``order'' instead selects all tracks containing an event overlapping the time interval $(t-\epsilon, t+\epsilon)$, and orders them based on their temporal match.
     
	 \begin{figure}[tb]
	 \centering % avoid the use of \begin{center}...\end{center} and use \centering instead (more compact)
     \includegraphics[width=\columnwidth]{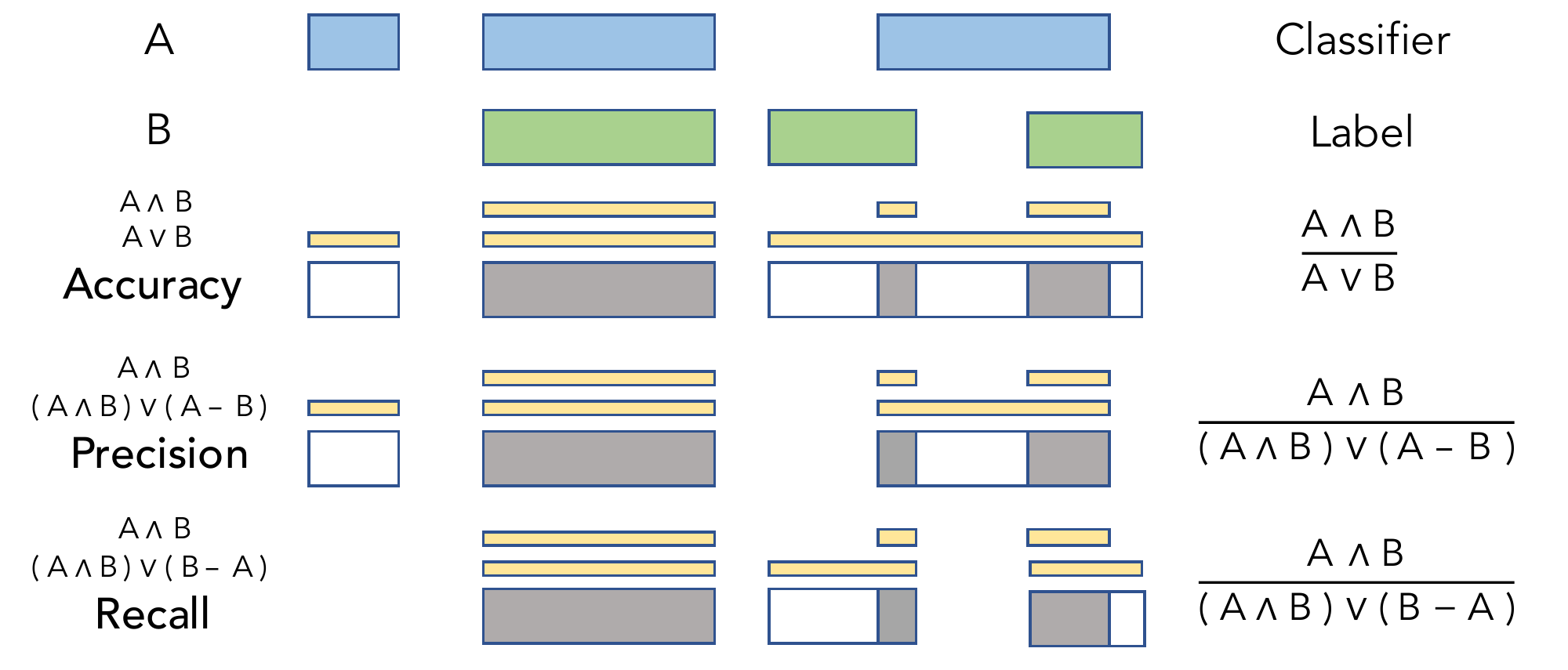}
     \caption{
     Common performance metrics computed through track algebra. $A$ is a
     classifier track and $B$ is a label track containing ground-truth
     labels. Accuracy, precision and recall scores can be visualized as 
     a ``container'' track with partial color fill. Mispredictions affecting 
     a specific performance metric are localized in the non-filled (blank) 
     regions of the track. \label{fig:metrics}}
    \end{figure}

    \begin{figure}[tb]
     \centering % avoid the use of \begin{center}...\end{center} and use \centering instead (more compact)
     \includegraphics[width=\columnwidth]{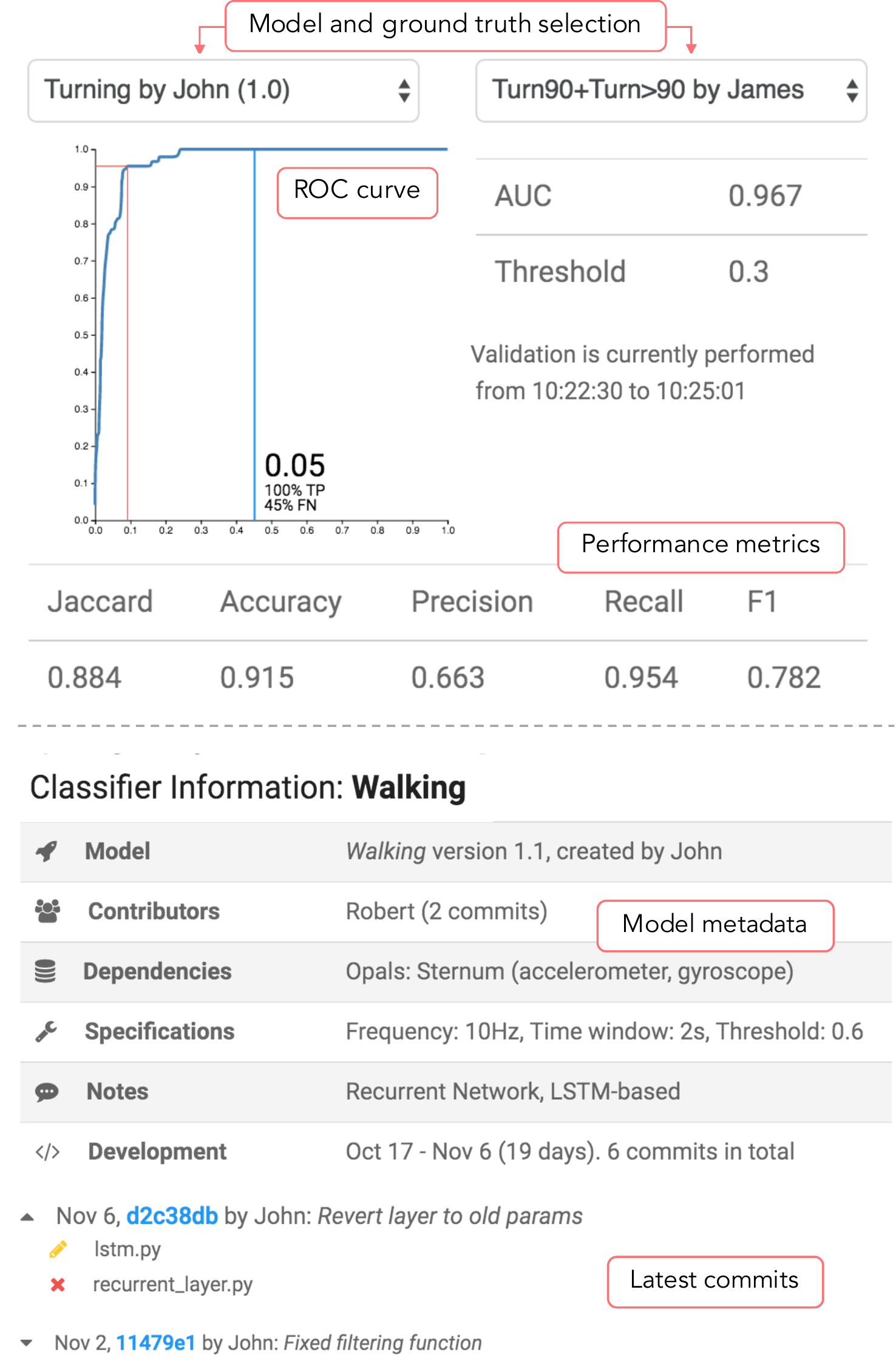}
     \caption{Performance metrics (top) and classifier information (bottom)
       modal windows. Xplorer features a modal window (top) to display
       different performance measures for a classifier track 
       % with respect to a label track, -- CD
       and includes an interactive ROC (Receiver Operating
       Characteristic) curve to help the user choose an adequate threshold for
       the selected model. The information modal (bottom) displays model
       metadata and summarizes commits and modifications performed on the code
       repository. \label{fig:validation}}
    \end{figure}
    
    \subsection{Classifier Validation}
    When a classifier track $A$ and a ground-truth label track $B$ are observed next
    to each other, it is intuitively clear that the performance of the
    predictive model depends on how extensively the blocks of each track are aligned with
    each other. Optimally, for each block in $A$ there should exist a block in $B$
    of equal length whose start and end points match those of $A$.
    However, misclassifications and other prediction-related errors may misalign or eliminate one of these two blocks.  
    
    A straightforward numerical way to quantify the visual overlap of two
    tracks is the Jaccard index, defined as the intersection over
    union of two series of values.  If we define the ``length'' of a track as
    the sum of the duration of all its blocks, the Jaccard index simply
    corresponds to the length of the intersection of two tracks ($A \wedge B$)
    divided by the length of their union ($A \vee B$). 
    
    Leveraging the definitions in Fig.~\ref{fig:track_algebra} lets us extend this example and express common performance metrics such as
    accuracy, precision and recall scores in terms of visual track algebra
    (Fig.~\ref{fig:metrics}). In particular, we can define a new type
    of track in which a set of ``empty'' boxes is defined by the denominator of
    the formulas in Fig.~\ref{fig:metrics} and is filled according to the
    numerator of the fraction. Non-filled regions of the track correspond to
    mispredictions affecting a specific performance metric, whereas filled
    regions correspond to the contributing correct predictions. While
    performance metrics are generally expressed as aggregate numerical values,
    our method also visualizes them spatially---enabling data scientists to
    understand \textit{how} these proportions are computed, to identify
    \textit{where} mispredictions are located in time, and to discern
    \textit{which} of them affect a specific metric of interest. For instance,
    the diagram in Fig.~\ref{fig:metrics} makes it clear that the
    value of the recall performance metric is unaffected by the first
    misprediction of classifier A.

Since performance metrics often depend on the threshold applied to the
continuous prediction of a classifier, we include in the Xplorer interface a
modal window displaying an interactive Receiver Operating Characteristic (ROC)
plot (Fig.~\ref{fig:validation}) with its related Area Under the Curve (AUC)
score---a threshold-independent performance metric. By dynamically changing
the threshold, the user can balance the number of true positives and false
negatives and observe how tracks and metrics are updated.
    
 %By sampling each track into a sequence of prediction values or binary labels, it is possible to compute different performance metrics common in data science such as accuracy score, precision, recall and F1 score.

    \begin{figure*}
    {
      \centering
    \includegraphics[width=\linewidth]{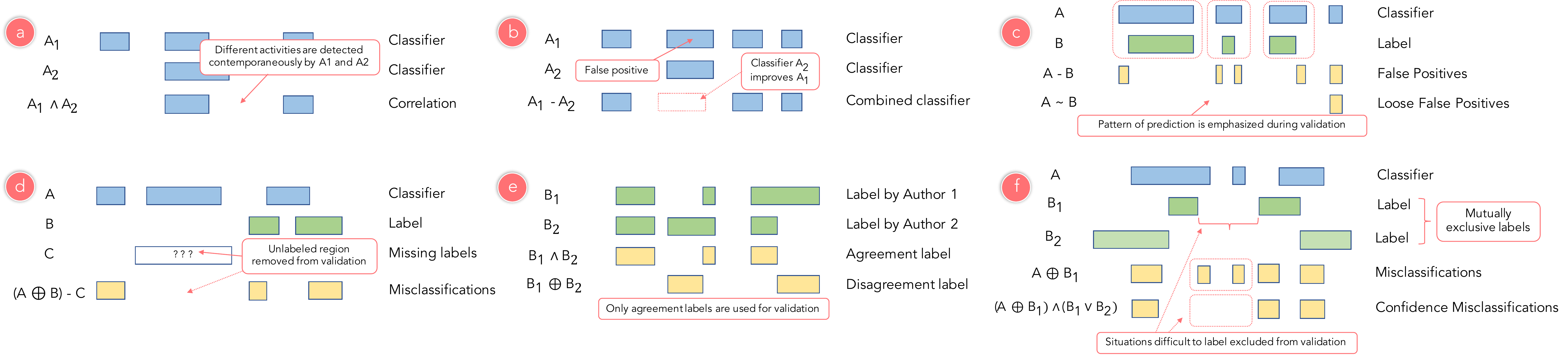}
      \caption{Advanced algebraic applications. Track algebra can also be
	applied among classifier tracks, e.g., to quantify their
	correlation (a) or to estimate their combined performance if they
	were executed together (b). Other relevant applications relate to
	obtaining a more legitimate performance estimate when dealing with
	missing labels (d), unclear motor events (f), or multiple sources of
	ground-truth information (e). The ``\textasciitilde'' operator is also
	introduced to emphasize event detection over precise temporal
	matching (c).  \label{fig:advanced}}
      }
    \end{figure*}
    
\subsection{Advanced Algebraic Combinations}
    
    So far we have primarily described how to apply our visual track algebra between a classifier track and label track containing ground-truth
    labels. In real-life scenarios, the development of predictive models can
    benefit from applying algebra on a wide variety of track combinations, even
    over more than two tracks at a time.
    
	It can be meaningful to apply our algebra over two classifier
	tracks. Computing the intersection of two classifiers
	(Fig.~\ref{fig:advanced}a) shows how greatly their event
	detections correlate. Alternatively, one can combine
	multiple classifiers to estimate how they would perform if executed
	together. As an example, if a predictive model has many
	false positives because the subject is performing a different activity,
	a second classifier that detects that activity
	and removes the outliers (Fig.~\ref{fig:advanced}b) could be useful.
    
    Often the generation of ground-truth information for motor activities entails
    multiple people manually annotating the same dataset. Since motor
    activities are perceived slightly differently by annotators, evaluating performance based on their agreement
    (Fig.~\ref{fig:advanced}e) is often useful. Track algebra can also be used to exclude from
    validation all cases in which ground-truth
    labels (Fig.~\ref{fig:advanced}d) could not be collected, thus guaranteeing a more legitimate
    performance estimate. Similarly, mutually exclusive labels can be combined
    to exclude unclear motor events (Fig.~\ref{fig:advanced}f). 
    
    Finally, we note that classification performance in activity detection is
    highly affected by how much predictions and labels overlap over time.
    Sometimes it is of interest to precisely matching the start and end time
    of each motor event, whereas in other instances event detection is sufficient. In
    the latter case, precise overlap of two tracks is not a reasonable
    performance metric. We thus introduce the ``\textasciitilde''
    operator, a variation of the subtraction operator that is not sensitive to
    the start and end of each detection (Fig.~\ref{fig:advanced}c).
    Applications of the advanced algebraic operations introduced here are
    discussed further  in the next section.

     \begin{figure*}
    {
      \centering
      \includegraphics[width=\linewidth]{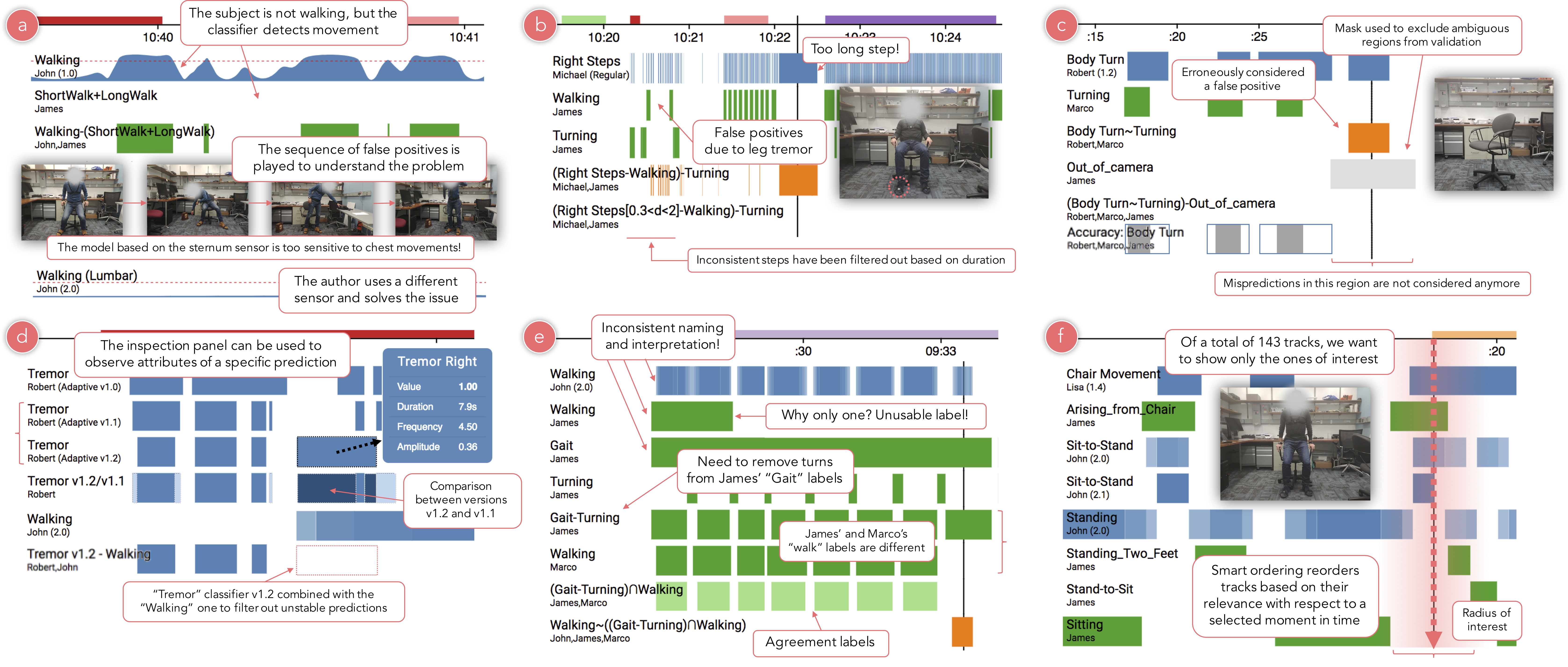}
      \caption{Sample screens from our use case: (a) combining track
	algebra and video playback to identify false positives; (b) using
	filtering to identify and remove tremor mispredictions; (c)
	removing regions of uncertainty from validation; (d) comparing classifier
	version and inspecting attributes; (e) handling inconsistent
	labeling with track algebra; (f) using smart ordering to identify
	tracks of interest. \label{fig:usecase}}
      }
    \end{figure*}

\section{Use Case: Detecting Motor Activities in Parkinson's Subjects}\label{sec:usecase}
% ---- USE CASE: DETECTING MOTOR ACTIVITIES OF PATIENTS WITH PARKINSON'S DISEASE ---- 
   
    %What it is BlueSky project \cite{Bluesky} aims at deploying predictive
    %models to automatically assess the symptoms of Parkinson's disease using
    %wearable sensors.
    We present here the application of our system to a project involving 
    predictive models to automatically detect specific motor
    activities by subjects affected by Parkinson's disease.
    %Xplorer and bluesky
    Xplorer was used as a companion tool over most of the project by a mixed team of
    fourteen data scientists, domains experts and business people.

    %Explain the data and preparation
    Six wearable IMU sensors were worn by 25 Parkinson's disease
    subjects over multiple sessions (visits) of about an hour's duration. The sensors measured
    accelerometer, gyroscope and magnetometer information at 128Hz and were
    placed on the wrists, feet, chest and back of the patients. During each
    visit, all subjects performed the same set of predefined tasks according
    to a single clinical protocol. Technicians not involved in the system development work handled recording sessions, labeling specific activities and time-stamping the task
    executions.  Sensor data, ground-truth labels and video files were
    all stored in a single database that all data scientists could access
    as they developed their predictive models.
    
    %Data scientists
    Data scientists developed their algorithms in Python, in accordance
    with the specifications of our analytics pipeline. Their code was pushed to
    a private GitHub repository and automatically loaded and executed by our
    system. The results of the computation were systematically made available
    to the team through the Xplorer user interface, with no additional
    input from the group members. Classification results were analyzed
    both individually and collaboratively during weekly team
    meetings.

\subsection{Classifier Development and Comparison}

Developing a predictive model is an iterative, trial-and-error process made up of analyzing a first set of classification results, tuning model
parameters and then recomputing predictions. Xplorer proved valuable to
data scientists in discovering insights as they analyzed classification
results.

\bfparhead{Video Playback and Track Algebra} The playback functionality, in
combination with the track algebra, proved to be fundamental in
quickly identifying mispredictions. For example, subtracting the ``Walk''
label track from the ``Walk'' classifier track and playing the resulting
track showed all cases in which the classifier did not
detect that the subject was walking (false negatives). By observing the video
and the task labels, data scientists realized that, since the model was using a sensor worn on the chest, it was incorrectly classifying movements such as
rising from the chair and buttoning a coat (Fig.~\ref{fig:usecase}a).  Similarly, the ``Step detector''
classifier track (based on shoe-worn sensors) showed false positives corresponding to
foot tremor (Fig.~\ref{fig:usecase}b), particularly common when subjects were sitting
with their legs crossed.  Exploiting these insights,
data scientists decided to re-train their classification models, either by using data from
different sensors or by including in the training set the activities that had
been misclassified.

\bfparhead{Model Debugging} Another widely used feature was the ability to
inspect information about each single prediction.  After noticing that
the two hand classifiers ``Pronation-supination'' and ``Tremor'' were biased by
the action of walking, data scientists were able to mouse over mispredicted
events and observe the attributes computed by their predictive models (Fig.~\ref{fig:usecase}d). In this
case, each prediction held numerical information about hand rotation angle,
hand rotation speed, tremor frequency and tremor amplitude. By analyzing these
attributes, data scientists were able to filter out movements happening at
specific frequencies associated with walking, thus making their model more
robust.
%Filtering
Similarly, the filtering function was used to check the validity of step
detections and revealed several events characterized by unexpectedly long
durations and unrealistic speeds (Fig.~\ref{fig:usecase}b). Isolating these events and playing their
associated video showed that the step-detection algorithm did not properly 
detect the foot-landing phase.
%Download
%Dynamical visualization and download of the raw data associated to a time period were used respectively to verify the correct behavior of algorithms and to precisely isolate sensor data for further offline analysis.

\bfparhead{Classifier Versioning} A key feature of Xplorer is the ability to compare the results not only of different classifiers, but also of different versions of the same predictive model. Not always updating a
model may yield better performance, and tracking changes is
important in understanding which modifications have led to an improvement or
should be reverted. Version comparison was typically used to tune model
parameters to balance the number of false positives and false negatives, in
particular for classifiers such as ``Sit2Stand'' and ``Tremor''
(Fig.~\ref{fig:usecase}d). 

\subsection{Label Accuracy and Validation}
While human-generated labels may seem a valid source of
ground-truth information, blindly assuming their completeness and correctness
can easily lead to wrong insights and inaccurate performance estimates. A mislabeled
time period, for instance, can erroneously reveal a false-positive or false-negative prediction, thus decreasing the classifier's
performance and creating a bias in interpreting its classification results.
%affecting the analysis of the causes of the (non-existing) misprediction.

%Track algebra for union of labels
\bfparhead{Partial Labeling} The first issue encountered in this project
was the discrepancy between the ground-truth labels provided by video
annotators (who were instructed using a pre-existing medical protocol) and
those requested or expected by data scientists (who were more aware of machine-
learning requirements for training). In many cases, such problems could be
easily solved by combining labels through track algebra. For instance, while
raters defined partial labels such as ``ShortWalk'' and ``LongWalk'' or
``Turn90'' and ``Turn>90'', data scientists could combine them in single
tracks (``ShortWalk+LongWalk'', ``Turn90+Turn>90'') and then use them in
evaluating the performance of their classifiers (Fig.~\ref{fig:detail}).

%Out of camera
\bfparhead{Missing Labels} Another problem involving manual video annotations
was a sporadic absence of labels, either due to distraction of the technician
or to the subject being off camera. In evaluating the performance of the
``Walking'' classifier through track algebra, the team found a percentage of
false positives much higher than expected. Playing the corresponding parts
of the video revealed that the subject had walked outside the camera
view, where the video annotator couldn't tell what actions were performed. Here, the annotator was asked to generate a new label track indicating all moments in which the patient was
off camera, and these were then excluded from the performance evaluation via track
algebraic subtraction (Fig. \ref{fig:usecase}c).

\bfparhead{Label Definition and Human Bias} In looking more minutely at mispredictions, the team observed that among the causes of lower
performance was the definition of motor activities themselves. For
instance, while the ``Walking'' classifier was trained to recognize any
horizontal movement involving foot motion, the annotator's definition of the
motor activity required a minimum of three steps by the patient in
the same direction. Because of this, the track algebraic difference showed a
large set of false positives associated to small movements, such as performing
few steps in reaching an object. Similarly, an
inconsistent definition of ``turn'' was used for generating labels. Typical questions in team meetings included
``Should a larger rotation of the chest be considered a turn, even if the legs
don't move?'', ``Which is the minimum angle of rotation that defines a turn?'',
and ``To what extent it is useful to consider such details for the purpose of
the project?''. Fig~\ref{fig:usecase}e shows how these inconsistencies in labeling were handled through track algebra.
%Human bias
Visually matching the alignment of predictions revealed human bias in the annotation of motor activities. For instance, it was
discovered that some ``walking'' labels were assigned \textit{before} the patient
actually started moving his feet. This mislabeling probably occurred because the video annotator already expected the intention to move of the subject, whose actions had to follow a standard medical protocol. During performance
evaluation, data scientists could limit these difficulties by considering
detection patterns through the ``\textasciitilde'' operator
(Fig~\ref{fig:advanced}c).
%Multiple raters
To mitigate this problem, the data scientist team asked
the technicians to provide multiple ground-truth labels by different video annotators for a given activity.
Thanks to Xplorer's track intersection functionality (Fig.~\ref{fig:advanced}e), data scientists could then run performance evaluations only in regions where the annotators agreed (Fig.~\ref{fig:usecase}e).

\subsection{Team Collaboration}
While individual data scientists used Xplorer for classifier
development, the visualization tool was also used collectively by the entire
team during weekly meetings. In particular, its visual output was projected on
a large screen or shared with remote participants via video conferencing. Every
week, the tool would be used to show progress on acquiring new
patient data, developing classifiers and generating new
ground-truth labels. The person acting as moderator also used Xplorer to 
review team progress and lead the conversation.

\bfparhead{Collaborative Interpretation} Due to its abstraction and simple
visuals, Xplorer proved to be an efficient medium for discussion among people
with very different backgrounds. Lack of familiarity with machine
learning or with the details of each predicted model was no impediment for team members in visually checking the alignment of tracks and further validating
them with the video.  Each data scientist could observe and give feedback
about models built by other developers without knowing all their
implementation details. Domain experts as well were able to share their
knowledge of Parkinson's disease and advise data scientists how best to
handle certain motor-activity events. Similarly, data scientists could
express through Xplorer the need for additional training samples or higher-
quality ground-truth labels.

\bfparhead{Decision Making} Xplorer was used managerially
to track the progress of the project with respect to deadlines and to assess
the quality of each predictive model before deployment. Human resources were dynamically allocated in order to compensate for
predictive models that showed weaknesses in Xplorer. For instance, using the tool
quickly led to a discussion of the quality and reliability of labels and suggested that
management consider hiring new personnel to annotate videos manually.
Similarly, false negatives shown in Xplorer were used to justify the
acquisition of new data from additional patients in order to have enough
samples for each Parkinson's phenotype. Consideration of classifier accuracy from the video
footage led to the exclusion of a subset of the sensors from the project scope, a decision that led data scientists to focus more on wrist and
lumbar sensors.

\section{Discussion}\label{sec:discussion}
% ---- DISCUSSION ---- 

\subsection{Improving Scalability at Multiple Levels}
Users can exploit five main data components in performing visual analysis of
classification results: (1) the actual results of the analysis, (2)
ground-truth labels, (3) multimedia footage, (4) raw sensor data and (5)
additional metadata information.  While (3) can be consumed via streaming and
(5) generally uses only a limited amount of memory, the number of instances of
(1), (2) and (3) can cause serious scalability issues that affect the
interactivity of the system. We resolve these issues by reducing the frequency
of (1) and (2) events through the compression algorithm in
Fig.~\ref{fig:compression}, and by dynamically loading (3) only for small
subsets of the dataset, based on user request.

%This allowed Xplorer to run smoothly using SVG only in the BlueSky use case. 
	
A different scalability limitation lies instead in the number of tracks
visualized on screen, which amounted to almost 100 in our use case scenario.
Too many classifier or label tracks displayed at the same time can make visual
search and comparison cumbersome for the user, who would prefer a small,
ordered set of tracks so as to achieve the desired visual insights. For this
reason, we combine the use of smart autocompletion and wildcard selectors in
the command line with a smart ordering functionality.

%Smart ordering
Smart track ordering addresses the situation in which the user is interested in
a particular moment in time and wants to visually compare all tracks containing
relevant information in that particular time window. Say, for instance, that
the user is focusing on the moment in which the subject is standing up from a
chair and wants to see tracks that detect events related to this action. Smart
ordering automatically reorders classifier and label tracks, bringing Arising
from chair, Sit-to-Stand, Sitting and Standing close to each other. Smart
ordering further remembers previous tracks of interest, so that if the user
focuses on a consecutive action in which the user sits back on the chair, the
track Stand-to-sit is ordered on top and the track Sit-to-stand (no longer
involved) remains visible.
	
%Wildcards are a useful alternative to track identifiers autocompletion
%in the command line. When having too many tracks, it is easy to forget their
%name or id and it is sometimes more convenient to identify them by their order
%of appearance (e.g. ``add 1 2'' adds the first two tracks). When using
%wildcards, priority is given to tracks based on their order of appearance,
%making it easier to combine the use of smart ordering with track algebra.

\begin{figure}[tb]
     \centering
     \includegraphics[width=\columnwidth]{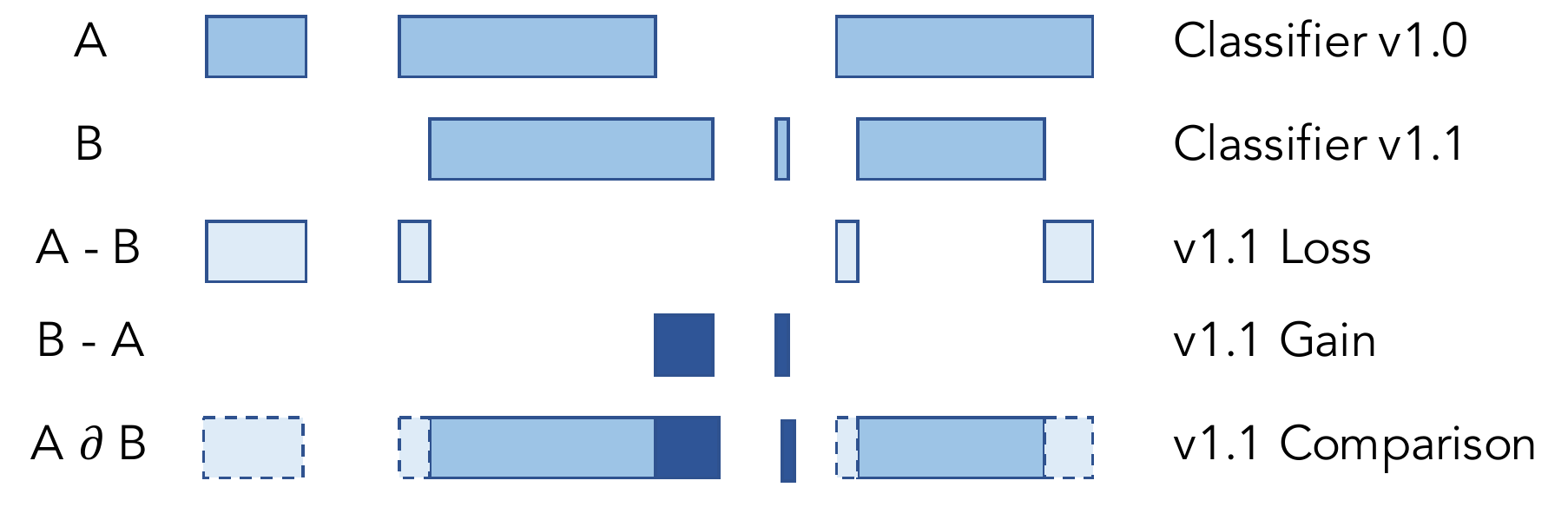}
     \caption{Classifier versioning. Track Xplorer supports a hybrid versioning
       based on the combination of classic version control systems (e.g.
       git) and a standardized model definition. Track algebra can also be used
       to visualize differences between classifier versions.\label{fig:versioning}}
\end{figure}    
    
\subsection{A First Step Towards Model Versioning}
% With the incredible performance and generalization possibilities offered by
% machine learning, research has more and more focused on creating systems to
% build better models. However, despite the increasing use of this technology
% across several domains, many aspects of the predictive modeling life cycle have
% not been properly addressed yet.
% ModelHub \cite{miao2017modelhub} tries to address the issues of model storage, versioning and sharing by introducing a new versioning system and a query language specific to the machine learning domain. Similarly, Van der Weide et al.\cite{van2017versioning} propose versioned pipelines as a custom solution, leveraging asynchronous deployment to optimize prediction throughput.
% Instead, our 
As machine learning technology matures, research has focused more
and more on creating systems to build models more effectively.
However, despite the fast expansion of machine-learning applications into new
domains, many aspects of the predictive modeling life cycle have not yet been
properly addressed.
%ModelHub \cite{miao2017modelhub} tries to address the issues of model storage, versioning and sharing by introducing a new versioning system and a query language specific to the machine learning domain. Similarly, Van der Weide et al.\cite{van2017versioning} propose versioned pipelines as a custom solution, leveraging asynchronous deployment to optimize prediction throughput.
%Instead, our 
Track Xplorer relies on
enforcing a standardized model definition that all data scientists commit
to use, but leaves them free to continue using \textit{git} as a versioning
system. In particular, Xplorer's analytics pipeline automatically extracts 
model metadata information from \textit{git} commits and presents it in the
front-end user interface (Fig.~\ref{fig:validation}, bottom) while comparing multiple versions of the same
classifier. Data scientists can use the track algebra to identify differences
between model versions (Fig.~\ref{fig:versioning}) and then inspect which parameters
and commits are associated to the best results. This way, data scientists can
easily go back to their code and revert or confirm modifications to their
model.

\begin{figure}[tb]
     \centering
     \includegraphics[width=\columnwidth]{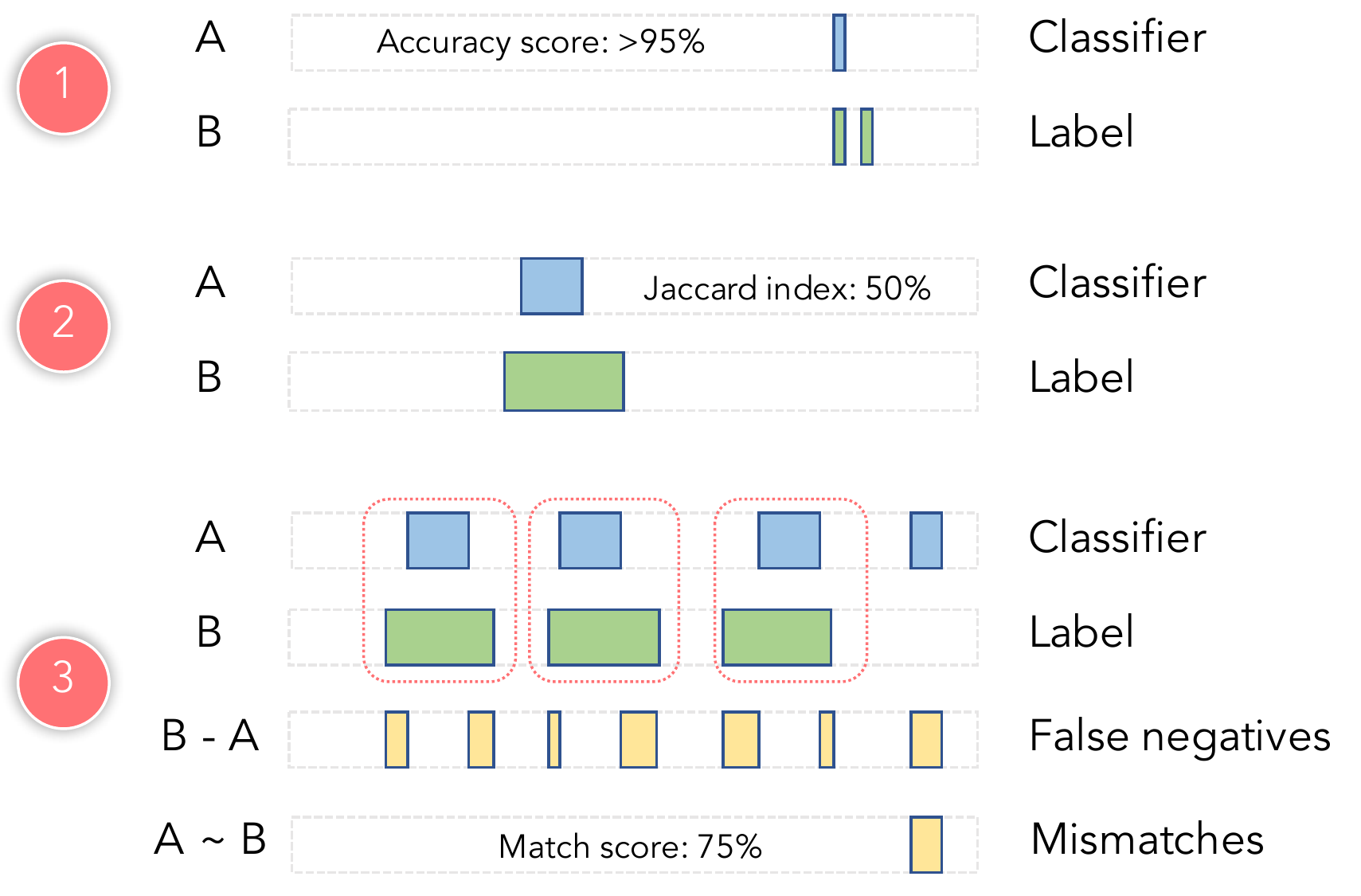}
     \caption{Enhanced performance analysis through visual track algebra. Track
       Xplorer lets data scientists reason about the tradeoffs of each
       performance metric, helping them decide which ones to adopt. 
       It also helps them gain insights for developing new, effective performance 
       metrics tailored to the classification task at hand.\label{fig:quality}} 
 \end{figure}
    
\subsection{Running Meaningful Performance Evaluations}
     As shown in Fig.~\ref{fig:metrics}, our track algebra also supports
     visualizing over time the contributions of predictions to the computation
     of common performance metrics. Key in evaluating the
     performance of a predictive model is choosing which quality metrics to
     adopt --- and the decision is often based on the
     desired outcome of the prediction. For instance, let's suppose we are
     interested in detecting the event of a subject standing from a seated
     position, but we are not concerned with detecting a precise time
     window for this movement. In case (1) of Fig.~\ref{fig:quality}, we
     would correctly predict one Sit-to-Stand event out of two, after a long
     period in which no events are detected. If we consider the fact that the
     classifier correctly predicted that no other events happened over that
     period, the accuracy metric would tell us that our model is
     performing extremely well (beyond 95\% correct predictions over time).
     However, for our purposes, this metric is extremely misleading, since the
     classifier missed half of the interesting events. Similarly, the Jaccard
     index computed in case (2) of Fig.~\ref{fig:quality} would suggest a 50\%
     overlap, but this is not very meaningful if we are interested in event
     detection only. The simple concepts of false positives and false negatives
    in fact assume subjective meanings in time-series data classifications.
     Fig.~\ref{fig:quality}(3) shows how the detection of four events would by
     definition generate seven sequences of mispredictions, with a very
     negative impact on most performance metrics. Xplorer enables data
     scientists to visually assess the suitability of a particular metric, but
     also encourages the definition of new performance estimators. For
     instance, in Fig.~\ref{fig:quality}(3) we could approximate the overall
     performance by applying the \textit{match} operator of our track algebra,
     which identifies only one mispredicted event. Visually, this leads us to
     define a new and more suitable empirical performance metric that
     simply corresponds to the number of correctly detected events (in
     this case, three out of four, i.e. 75\%).

 \begin{figure}[tb]
     \centering
     \includegraphics[width=\columnwidth]{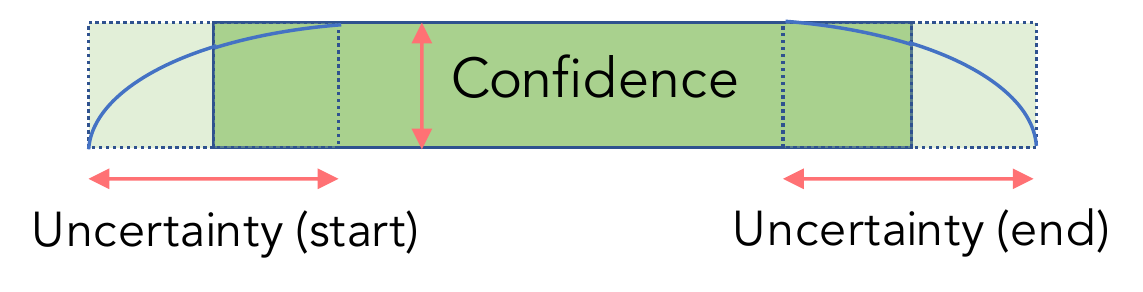}
     \caption{A probabilistic approach to ground truth information. Labels
       could be better modeled as a function over time to take into account the
       confidence on the event and the uncertainty in precisely identifying its
       start and end timepoints.\label{fig:probability}
       \vspace{-2em}}
 \end{figure}     
     
\subsection{A New Model for Activity Labeling}
    Another issue affecting performance evaluation is the quality of ground-
    truth labels, which we treated simply as actions defined through a start
    and end moment in time. As discussed in our use case, video
    annotators cannot always find precise time boundaries for a motor activity
    and at times are not even sure if an action should be labeled at all---with 
    predictably affecting the subsequent validation process. It may be
    useful instead to approach ground-truth labels probabilistically,
    similarly to what is currently done for classifiers. For
    instance, a ground-truth label could be defined as a function whose value
    depends on the confidence of the video annotator, with time boundaries
    modeled so as to take into account the uncertainty in determining the exact
    moment at which the action started or ended (Fig.~\ref{fig:probability}).This would make it possible to devise new performance metrics and define a continuous version of our
    current track algebra, with broader possibilities for model
    analysis and manipulation.

\section{Conclusion}\label{sec:conclusion}
% ---- CONCLUSION ----  
We introduce Track Xplorer, a system for interactive visual analysis of
predictions of classifiers modeled to detect events in temporal sensor data.
Our system enables the user to visually and quantitatively analyze and
compare results from multiple classification models, improving the model
development and debugging experience of data scientists. Track Xplorer
couples contextual information such as ground-truth labels, expert
annotations and event videos together with track visualizations of
predictions through interaction and visual encoding, thereby empowering users
of diverse backgrounds to better interpret, debug, and enhance classifier
performance.  We also introduce an extensible visual algebra over track
representations that supports composable and rigorous performance comparison
and analysis.
   
We demonstrate the usefulness of our tool through its application in a
collaborative project for classifying motor activity patterns to score the
degree of disease progression among Parkinson's disease patients. Track
Xplorer enables the project team members to identify early on possible
systemic errors in the data, reason about and find the causes of
misclassifications, and effectively compare the results of different
classifiers, thus improving classification performance by selecting better
models and parameters.  

Research into systems for machine learning  model development and management
is nascent but burgeoning. These systems would benefit from integrating data
visualization and visual analytics as a first-class citizen for improved user
experience. Track Xplorer is the first system that enables visual and
quantitative analysis, comparison and tracking of temporal multisensor data
classifications across models.

\section{Acknowledgments}\label{sec:ack}
We thank the BlueSky project members at IBM Research for their feedback 
on Track Xplorer and earlier drafts of this paper.

\bibliographystyle{eg-alpha-doi}
\balance
\bibliography{paper}

\newcommand{\etalchar}[1]{$^{#1}$}
\begin{thebibliography}{\uppercase{MLMdO{\etalchar{*}}13}}

\bibitem[ACD{\etalchar{*}}15]{modeltracker2015chi}
\textsc{Amershi S., Chickering M., Drucker S.~M., Lee B., Simard P., Suh J.}:
\newblock {ModelTracker}: Redesigning performance analysis tools for machine
  learning.
\newblock In \emph{Proc. CHI} (2015), ACM, pp.~337--346.

\bibitem[{Ado}17]{premiere}
\textsc{{Adobe Corporation}}:
\newblock {Premiere Pro}.
\newblock \url{http://www.adobe.com/products/premiere.html}, 2017.
\newblock Accessed: 2017-07-31.

\bibitem[AHH{\etalchar{*}}14]{wheel2014tvcg}
\textsc{Alsallakh B., Hanbury A., Hauser H., Miksch S., Rauber A.}:
\newblock Visual methods for analyzing probabilistic classification data.
\newblock \emph{{IEEE TVCG} 20}, 12 (2014), 1703--1712.

\bibitem[All83]{allen1983cam}
\textsc{Allen J.~F.}:
\newblock Maintaining knowledge about temporal intervals.
\newblock \emph{Commun. ACM 26}, 11 (1983), 832--843.

\bibitem[{App}17]{imovie}
\textsc{{Apple Corporation}}:
\newblock {iMovie}.
\newblock \url{https://www.apple.com/imovie/}, 2017.
\newblock Accessed: 2017-07-31.

\bibitem[BCH{\etalchar{*}}15]{datahub}
\textsc{Bhattacherjee S., Chavan A., Huang S., Deshpande A., Parameswaran A.}:
\newblock Principles of dataset versioning: Exploring the recreation/storage
  tradeoff.
\newblock \emph{Proceedings of the VLDB Endowment 8}, 12 (2015), 1346--1357.

\bibitem[BSM04]{bade2004chi}
\textsc{Bade R., Schlechtweg S., Miksch S.}:
\newblock Connecting time-oriented data and information to a coherent
  interactive visualization.
\newblock In \emph{Procs. CHI} (2004), ACM, pp.~105--112.

\bibitem[CFS{\etalchar{*}}06]{callahan2006vistrails}
\textsc{Callahan S.~P., Freire J., Santos E., Scheidegger C.~E., Silva C.~T.,
  Vo H.~T.}:
\newblock Vistrails: visualization meets data management.
\newblock In \emph{Proceedings of the 2006 ACM SIGMOD international conference
  on Management of data} (2006), ACM, pp.~745--747.

\bibitem[CvW18]{cappers2018exploring}
\textsc{Cappers B.~C., van Wijk J.~J.}:
\newblock Exploring multivariate event sequences using rules, aggregations, and
  selections.
\newblock \emph{IEEE TVCG (Proc. InfoVis)} (2018).

\bibitem[DHHH13]{Demiralp2013}
\textsc{Demiralp C., Hayden E., Hammerbacher J., Heer J.}:
\newblock invis: Exploring high-dimensional rna sequences from in vitro
  selection.
\newblock In \emph{Proc. BioVis} (2013), IEEE, pp.~1--8.

\bibitem[FKSS06]{fails2006vast}
\textsc{Fails J., Karlson A., Shahamat L., Shneiderman B.}:
\newblock A visual interface for multivariate temporal data: Finding patterns
  of events across multiple histories.
\newblock In \emph{Procs. VAST} (2006), pp.~167--174.

\bibitem[FWBB10]{savant:bioinf10}
\textsc{Fiume M., Williams V., Brook A., Brudno M.}:
\newblock Savant: genome browser for high-throughput sequencing data.
\newblock \emph{Bioinformatics 26}, 16 (2010), 1938--1944.

\bibitem[GS14]{gotz2014decisionflow}
\textsc{Gotz D., Stavropoulos H.}:
\newblock Decisionflow: Visual analytics for high-dimensional temporal event
  sequence data.
\newblock \emph{IEEE transactions on visualization and computer graphics 20},
  12 (2014), 1783--1792.

\bibitem[HOB94]{timelines1994gi}
\textsc{Harrison B.~L., Owen R., Baecker R.~M.}:
\newblock Timelines: An interactive system for the collection and visualization
  of temporal data.
\newblock In \emph{Procs. Graphics Interface} (1994), pp.~141--148.

\bibitem[JS99]{jensen1999review}
\textsc{Jensen C.~S., Snodgrass R.~T.}:
\newblock Temporal data management.
\newblock \emph{{IEEE TKDE} 11}, 1 (1999), 36--44.

\bibitem[JS09]{querymarvel2009vlhcc}
\textsc{Jin J., Szekely P.}:
\newblock Querymarvel: A visual query language for temporal patterns using
  comic strips.
\newblock In \emph{Procs. {VL/HCC}} (Sept 2009), pp.~207--214.

\bibitem[JS10]{querymarvel2010vast}
\textsc{Jin J., Szekely P.}:
\newblock Interactive querying of temporal data using a comic strip metaphor.
\newblock In \emph{Procs. {VAST}} (2010), pp.~163--170.

\bibitem[Kar94]{karam1994sigsoft}
\textsc{Karam G.~M.}:
\newblock Visualization using timelines.
\newblock In \emph{ACM SIGSOFT} (1994), pp.~125--137.

\bibitem[KSF{\etalchar{*}}02]{ucscbrowser2002gr}
\textsc{Kent W.~J., Sugnet C.~W., Furey T.~S., Roskin K.~M., Pringle T.~H.,
  Zahler A.~M., Haussler D.}:
\newblock The human genome browser at ucsc.
\newblock \emph{Genome Research 12}, 6 (2002), 996--1006.

\bibitem[MCD17]{miao2017provdb}
\textsc{Miao H., Chavan A., Deshpande A.}:
\newblock Provdb: Lifecycle management of collaborative analysis workflows.
\newblock In \emph{HILDA@ SIGMOD} (2017), pp.~7--1.

\bibitem[MLDD17]{miao2017modelhub}
\textsc{Miao H., Li A., Davis L.~S., Deshpande A.}:
\newblock Modelhub: Deep learning lifecycle management.
\newblock In \emph{IEEE 33rd International Conference on Data Engineering
  (ICDE)} (2017), pp.~1393--1394.

\bibitem[MLMdO{\etalchar{*}}13]{monroe2013chi}
\textsc{Monroe M., Lan R., Morales~del Olmo J., Shneiderman B., Plaisant C.,
  Millstein J.}:
\newblock The challenges of specifying intervals and absences in temporal
  queries: A graphical language approach.
\newblock In \emph{Procs. CHI} (2013), pp.~2349--2358.

\bibitem[PMR{\etalchar{*}}96]{lifelines1996chi}
\textsc{Plaisant C., Milash B., Rose A., Widoff S., Shneiderman B.}:
\newblock Lifelines: visualizing personal histories.
\newblock In \emph{Proceedings of the SIGCHI conference on Human factors in
  computing systems} (1996), ACM, pp.~221--227.

\bibitem[RAL{\etalchar{*}}17]{squares2017tvcg}
\textsc{Ren D., Amershi S., Lee B., Suh J., Williams J.~D.}:
\newblock Squares: Supporting interactive performance analysis for multiclass
  classifiers.
\newblock \emph{{IEEE TVCG} 23} (2017), 61--70.

\bibitem[RBER17]{ratner2017snorkel}
\textsc{Ratner A.~J., Bach S.~H., Ehrenberg H.~R., R{\'e} C.}:
\newblock Snorkel: Fast training set generation for information extraction.
\newblock In \emph{Proceedings of the 2017 ACM International Conference on
  Management of Data} (2017), ACM, pp.~1683--1686.

\bibitem[RDSW{\etalchar{*}}16]{ratner2016data}
\textsc{Ratner A.~J., De~Sa C.~M., Wu S., Selsam D., R{\'e} C.}:
\newblock Data programming: Creating large training sets, quickly.
\newblock In \emph{Advances in Neural Information Processing Systems} (2016),
  pp.~3567--3575.

\bibitem[RTW{\etalchar{*}}11]{igv:natureb11}
\textsc{Robinson J.~T., Thorvaldsdottir H., Winckler W., Guttman M., Lander
  E.~S., Getz G., Mesirov J.~P.}:
\newblock Integrative genomics viewer.
\newblock \emph{Nature Biotechnology 29}, 1 (2011), 24--26.

\bibitem[SAA{\etalchar{*}}94]{snodgrass1994tsql}
\textsc{Snodgrass R.~T., Ahn I., Ariav G., Batory D., Clifford J., Dyreson
  C.~E., Elmasri R., Grandi F., Jensen C.~S., K\"{a}fer W., Kline N., Kulkarni
  K., Leung T. Y.~C., Lorentzos N., Roddick J.~F., Segev A., Soo M.~D., Sripada
  S.~M.}:
\newblock Tsql2 language specification.
\newblock \emph{SIGMOD Rec. 23}, 1 (1994), 65--86.

\bibitem[SGM{\etalchar{*}}04]{ensembl2004gr}
\textsc{Stalker J., Gibbins B., Meidl P., Smith J., Spooner W., Hotz H.-R., Cox
  A.~V.}:
\newblock The ensembl web site: Mechanics of a genome browser.
\newblock \emph{Genome Research 14}, 5 (2004), 951--955.

\bibitem[SHG{\etalchar{*}}15]{sculley2015hidden}
\textsc{Sculley D., Holt G., Golovin D., Davydov E., Phillips T., Ebner D.,
  Chaudhary V., Young M., Crespo J.-F., Dennison D.}:
\newblock Hidden technical debt in machine learning systems.
\newblock In \emph{Advances in Neural Information Processing Systems} (2015),
  pp.~2503--2511.

\bibitem[SMWH17]{vegalite2017infovis}
\textsc{Satyanarayan A., Moritz D., Wongsuphasawat K., Heer J.}:
\newblock Vega-lite: A grammar of interactive graphics.
\newblock \emph{IEEE Transactions on Visualization and Computer Graphics 23}, 1
  (2017), 341--350.

\bibitem[STH02]{Stolte2002}
\textsc{Stolte C., Tang D., Hanrahan P.}:
\newblock {Polaris: a system for query, analysis, and visualization of
  multidimensional relational databases}.
\newblock \emph{{IEEE TVCG} 8}, 1 (2002), 1--14.

\bibitem[vdWPS{\etalchar{*}}17]{van2017versioning}
\textsc{van~der Weide T., Papadopoulos D., Smirnov O., Zielinski M., van
  Kasteren T.}:
\newblock Versioning for end-to-end machine learning pipelines.
\newblock In \emph{Proc. the 1st Workshop on Data Management for End-to-End
  Machine Learning} (2017).

\bibitem[VSL{\etalchar{*}}16]{vartak2016modeldb}
\textsc{Vartak M., Subramanyam H., Lee W.-E., Viswanathan S., Husnoo S., Madden
  S., Zaharia M.}:
\newblock {ModelDB}: a system for machine learning model management.
\newblock In \emph{Proc. HILDA} (2016).

\bibitem[WGGP{\etalchar{*}}]{lifeflow2011chi}
\textsc{Wongsuphasawat K., Guerra~G\'{o}mez J.~A., Plaisant C., Wang T.~D.,
  Taieb-Maimon M., Shneiderman B.}:
\newblock Lifeflow: Visualizing an overview of event sequences.
\newblock In \emph{Procs. CHI}, pp.~1747--1756.

\bibitem[Wil05]{Wilkinson_2005}
\textsc{Wilkinson L.}:
\newblock \emph{The Grammar of Graphics (Statistics and Computing)}.
\newblock Springer-Verlag, 2005.

\bibitem[ZDFD15]{zgraggen2015chi}
\textsc{Zgraggen E., Drucker S.~M., Fisher D., DeLine R.}:
\newblock (s$|$qu)eries: Visual regular expressions for querying and exploring
  event sequences.
\newblock In \emph{Procs. CHI} (2015), ACM.

\end{thebibliography}

\end{document}